\documentclass[reprint, showpacs,preprintnumbers,  amsmath,amssymb, aps]{revtex4-1}

\usepackage{graphicx}
\usepackage{dcolumn}
\usepackage{bm}

\newcommand\redout{\bgroup\markoverwith{\textcolor{red}{\rule[.5ex]{2pt}{0.4pt}}}\ULon}

\usepackage[colorlinks=true,citecolor=blue]{hyperref}
\hypersetup{colorlinks=true,citecolor=blue,linkcolor=red,urlcolor=blue}

\begin{document}
\title{Theory of strain in single-layer transition metal dichalcogenides}
\author{Habib Rostami}
\email{habib.rostami@nano.cnr.it}
\affiliation{NEST, Istituto Nanoscienze-CNR and Scuola Normale Superiore, I-56126 Pisa, Italy}
\affiliation{School of Physics, Institute for Research in Fundamental Sciences (IPM), Tehran 19395-5531, Iran}
\author{Rafael Rold\'an}
\affiliation {IMDEA Nanociencia Calle de Faraday, 9, Cantoblanco, 28049, Madrid, Spain}
\author{Emmanuele Cappelluti}
\affiliation{ISC-CNR, Deptarment of Physics, University of Rome ``La
  Sapienza'', P.le A, Moro 2, 00185 Rome, Italy}
\author{Reza Asgari}
\affiliation{School of Physics, Institute for Research in Fundamental Sciences (IPM), Tehran 19395-5531, Iran}
\author{Francisco Guinea}
\affiliation {IMDEA Nanociencia Calle de Faraday, 9, Cantoblanco, 28049, Madrid, Spain}
\affiliation{School of Physics and Astronomy, University of Manchester, Oxford Road, Manchester M13 9PL, UK}

\date{\today}

\begin{abstract}

Strain engineering has emerged as a powerful tool to modify the optical and electronic properties of two-dimensional crystals. Here we perform a systematic study of strained semiconducting transition metal dichalcogenides. The effect of strain is considered within a full Slater-Koster tight-binding model, which provides us with the band structure in the whole Brillouin zone. From this, we derive an effective low-energy model valid around the K point of the BZ, which includes terms up to second order in momentum and strain. For a generic profile of strain, we show that the solutions for this model can be expressed in terms of the harmonic oscillator and double quantum well models, for the valence and conduction bands respectively. We further study the shift of the position of the electron and hole band edges due to uniform strain. Finally, we discuss the importance of spin-strain coupling in these 2D semiconducting materials. 

\end{abstract}

\maketitle

\section{Introduction}

The outstanding stretchability of the new families of 2D crystals makes them excellent candidates for their use in strain engineering \cite{RG15}. This opens the possibility to fabricate nanodevices in which the optical and electronic properties are tunable by controlled application of external strain. Single layer of semiconductor transition metal dichalcogenides (TMDs) with the form $MX_2$ (where $M$=Mo, W is a transition metal and $X$=S, she is a chalcogen atom) can sustain large amounts of strain  before rupture of the membrane. For these materials, a direct-to-indirect band gap transition is expected for uniaxial/biaxial tensile strain of the order of $\sim 2-3\%$ \cite{WY14,FL12}, and a semiconducting-to-metal transition has been predicted for $10-15\%$ of tensile biaxial strain~\cite{SS12,GH13}.

Like graphene, low-energy excitations of insulating $MX_2$ are mainly localized close to the two inequivalent points K and K$'$, also denoted as ``valleys'', paving the way for the possibility of {\em valleytronics}, namely convey the information in the valley degree of freedom. The peculiarities of these materials suggest also possible ways to manipulate the valley-bit. The effect of strain on standard silicon semiconductor physics is known to lead to an enhancement of the electron and hole mobilities, and to the valley-degeneracy breaking~\cite{BS50}. However, silicon devices cannot sustain strains larger than $\sim 1.5\%$, whereas single layer TMDs support strength deformations higher than 10\%~\cite{BBK11,CR12}. The strong spin-orbit coupling (SOC) indeed yields a different spin-polarization of the valence band. Therefore, several degrees of freedom are strongly entangled in TMDs~\cite{XH14,RO14}. Tuning the spin-orbit coupling of mechanical deformation has been explored in conventional GaAs based semiconductors and quantum wells where a linear strain dependence is found in this coupling \cite{JW05,SA06,HW07}. Indeed very recently, a coupling between single electron spins and the motion of mechanical resonators based on crystal strain has been reported experimentally \cite{TM14}.
Therefore, controlling and tailoring their properties, at the applicant as well as at the theoretical level, represents thus the current challenge for a wide community of scientists.

In the last years, graphene has become the natural platform to test strain engineering physics.  One of the main theoretical study of deformed graphene-based material was done by Kane and Mele~\cite{Kane97} who used a tight-binding model to study the effect of long wavelength deformations on the low-energy electronic structure of carbon nanotubes. They showed that the effects of the tubule geometrical features and symmetry on its electronic structure are included through an effective vector potential. Such gauge field has been also predicted by Suzuura and Ando \cite{SA02} in the context of electron-phonon scattering in carbon nanotubes, and a group symmetry based survey has been done by Ma\~nes \cite{Manes07}. For specific profiles of strain, it was predicted theoretically~\cite{GKG10} and then proved experimentally~\cite{LC10,Lu12,HM12} that pseudo Landau level quantization corresponding to strong effective magnetic fields can be realized in graphene. Moreover, this kind of pseudomagnetic field is also observed in an artificial molecular graphene assembled by atomic manipulation of carbon monoxide molecules over a conventional two-dimensional electron system on a copper surface \cite{Gomes12}.

Strain engineering methods have been applied to other 2D crystals, and recently the possibility to tune the band gap with strain has been experimentally proven for MoS$_2$ \cite{CB13,HL13,CS13,ZU13} and WS$_2$~\cite{WY15,DC13,TM13}. Moreover, spatially modulated biaxial tensile strain has been applied to single layer MoS$_2$, leading to the realization of an optoelectronic crystal consisting of {\it artificial atoms}, due to the spatial modulation of the band gap in the sample \cite{Li2015}. Piezoelectricity and piezoresistivity effects have been recently reported for single layer and multi-layer MoS$_2$~\cite{Wu2014,MK15}. Therefore, there is a need for a deeper understanding of the effect of external non-uniform strain on the physical properties of semiconductor TMDs.

Here, we theoretically investigate the effect of strain on the electronic structure of a monolayer $MX_2$. 
Our main focus will be to study the effect of inhomogeneous strain on the low-energy physics of the system.
We start by considering a Slater-Koster tight-binding model which contains the relevant orbital character in the valence and conduction band, originated from $d_{3z^2-r^2}$, $d_{xy}$ and $d_{x^2-y^2}$ orbitals of the $M$ metal atom, and $p_x$, $p_y$ and $p_z$ orbitals of the chalcogen atom $X$~\cite{CG13,RO14}. Strain is considered in this model by means of a modification of the corresponding hopping terms~\cite{CS13}. From that model, we use the L\"{o}wdin partitioning method~\cite{W03} to obtain an analytical two-bands $\bf k\cdot p$ Hamiltonian valid in the vicinity of the K points of the BZ. This model differs on the previously used Dirac-like Hamiltonian~\cite{XY12,COG14} in the fact that we include terms up to second order in momentum and strain, which are needed to capture some of the main features of the $MX_2$ electronic band dispersion at low energies. One of the consequences is that strain leads to the appearance of not only one pseudo gauge field in the theory, but also to the existence of several pseudo vector potentials that couples in the relevant terms in the low-energy theory for $MX_2$, and which are absent in the well known case of strained graphene~\cite{VKG10,AV15}.

We further apply our model Hamiltonian to the case of arc-shaped deformation, and find that the corresponding solutions are well described in terms of the harmonic oscillator and double quantum well models, for the valence and conduction bands, respectively. Finally, we study the strain-induced change of the valence and conduction band edges, as well as the coupling between spins and strain on this family of TMDs.

The paper is organized as follows. In Sec. \ref{Sec:TB} we consider the effect of strain within a full Slater-Koster tight-binding model. From it, a low-energy $\bf k\cdot p$ Hamiltonian is derived in Sec. \ref{Sec:k.p}, and the impact of pseudo-vector fields on the electronic properties is discussed in Sec. \ref{Sec:Pseudo}. In Sec. \ref{Sec:SingleBand} we do an analytical analysis of the low-energy electronic spectrum of single layer TMDs using single band pictures.  In Secs. \ref{Sec:Valley} and \ref{Sec:Spin} we discuss the effect of strain on the position of the valence and conduction band edges, and spin-orbit coupling, respectively. Our main results are summarized in Sec. \ref{Sec:Summary}.

\section{Tight-binding model for strained TMDs }\label{Sec:TB}
Monolayer MoS$_2$ is a direct band gap semiconductor with the gap placed at the K and K' points of the hexagonal BZ. Ab-initio calculations show that there are two additional secondary extrema: a local maximum of the valence band at the $\Gamma$ point, and a local minimum of the conduction band at approximately at the Q point, midway between $\Gamma$ and K point~\cite{SW10}. These features, which are not relevant to the main optical properties of the system, might play an important role in transport properties \cite{SD13,KF13}. The low-energy physics of monolayer MoS$_2$ around the K and K' points was first described by a simple massive Dirac Hamiltonian \cite{XY12}. More accurate approximations have been developed later, as tight-binding methods \cite{RMA13,LX13,CG13,RL15} and ${\bf k\cdot p}$ approximations \cite{RMA13,KF15} which goes beyond the massive Dirac model, and account for the presence of trigonal warping and diagonal quadratic terms in momentum. In this section, we describe the TB  theory that will be used as starting point to consider strain effects on the electronic band structure of MoS$_2$. \par
\begin{figure}
\includegraphics[width=0.7\linewidth]{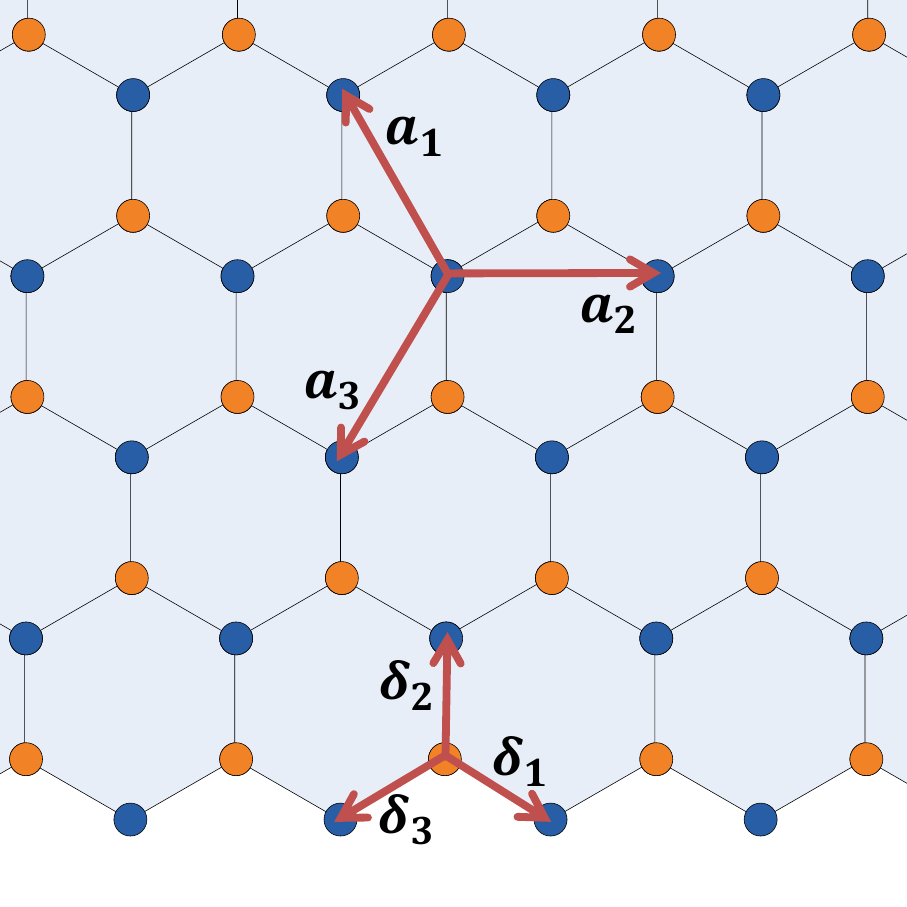}
\caption{(Color online) A top view schematic of monolayer MoS$_2$ lattice structure. Blue (Orange) circles indicate Mo (S) atoms. The nearest neighbors ($\delta_i$) and the next nearest neighbors ($a_i$) vector have been shown in the figure.} \label{Fig:Lattice}
\end{figure}


The main features of the band structure of monolayer MoS$_2$ in the whole BZ are well captured by the Slater-Koster TB model of Ref. [\onlinecite{CG13}] which includes 11 bands corresponding to the $d$ orbitals of the metal atom
and to the $p$ orbitals of the chalcogen atoms. Remarkably, the relevant physics of monolayer MoS$_2$ around the gap is covered by a smaller subspace, which can be obtained by performing an appropriate unitary transformation that transform the $p$ orbitals of the top and bottom S layers into their symmetric and antisymmetric combinations with respect to the $z$-axis. For the single-layer case, the resulting 11-band model
can be decoupled in 6 bands with even symmetry with respect to
$z \rightarrow -z$ inversion, and 5 bands with odd symmetry.
Low-energy excitations belong exclusively to the first block,
so that we will disregard the other states with different symmetry.
Local spin-orbit interaction can be as well included
in a suitable way~\cite{RO14}. Diagonal terms $\propto L_z S_z$
appear here to be dominant, so that in good approximation each spin
sector can be dealt  with separately~\cite{RO14}.
Using the compact notation of Ref. [\onlinecite{CG13}], we can consider 
the reduced Hilbert space:
\begin{equation}
\vec{\psi}
=
\left(d_{3z^2-r^2},~d_{x^2-y^2},~d_{xy},p_{x}^S,~p_{y}^S,~p_{z}^A\right)
\label{Eq:BaseE}
\end{equation}
where the \emph{S} and \emph{A} superscripts of the \emph{p}-orbitals refer to symmetric and antisymmetric combinations $p_i^{S}=1/\sqrt{2}(p_i^t+p_i^b)$ and $p_i^A=1/\sqrt{2}(p_i^t-p_i^b)$, the index $i$ runs over the spatial
directions $i=x,y,z$, and the labels $t$ and $b$ refer to the top and bottom sulfur planes, respectively. A top view of the crystal lattice of MoS$_2$ is sketched in Fig. \ref{Fig:Lattice}. The TB Hamiltonian defined by the base (\ref{Eq:BaseE}),
including the local spin-orbit coupling
can be expressed in the real space as
\begin{eqnarray}
\label{Eq:H-Real}
H
&=&
\sum_{i,\mu\nu}
\epsilon_{\mu,\nu} c^\dagger_{i,\mu}c_{i,\nu}
+
\sum_{ ij,\mu\nu}
{[t_{ij,\mu\nu} c^\dagger_{i,\mu}c_{j,\nu}+{\rm H.c.}]},
\end{eqnarray}
where $c^\dagger_{i,\mu}$ creates an electron in the unit cell $i$
in the atomic orbital labelled by $\mu=1,\ldots,6$ belonging to the Hilbert space
defined in Eq. (\ref{Eq:BaseE}). The Hamiltonian acquires a more compact form once written in the $k$-space:
\begin{eqnarray}\label{Eq:H-k}
H&=&\begin{pmatrix}H_{MM}&&H_{MX}\\{H_{MX}}^\dagger&&H_{XX}\end{pmatrix},\nonumber\\
H_{MM}&=&\epsilon_{M}+2\sum_{i=1,2,3}{t^{MM}_i\cos \left({\bf k}\cdot{\bf a}_i\right)},\nonumber\\
H_{XX}&=&\epsilon_{X}+2\sum_{i=1,2,3}{t^{XX}_i\cos\left( {\bf k}\cdot{\bf a}_i\right)},\nonumber\\
H_{MX}&=&\sum_{i=1,2,3}{t^{MX}_i e^{-i {\bf k}\cdot{\bm \delta}_i}},
\end{eqnarray}
where the nearest (${\bm \delta}_i$) and the next nearest (${\bf a}_i$) neighbour vectors are shown in Fig. \ref{Fig:Lattice}. All the hopping terms $t_{ij,\mu\nu}$ have been evaluated within a
Slater-Koster scheme~\cite{CG13,CS13,RO14,RA15}. For the sake of simplicity, we provide with the different contributions in Appendix \ref{App:TB}. An appropriate set of Slater-Koster parameters for MoS$_2$ is given in Table \ref{Tab:t-fit}.

\begin{table}[t]
\begin{tabular}{lclcr}
\hline
\hline
   Crystal Fields & \hspace{0.5truecm} &$\epsilon_0$ & \hspace{0.5truecm} &  -1.094 \\
                          & \hspace{0.5truecm} &$\epsilon_2$ & \hspace{0.5truecm} &  -1.512 \\
                          & \hspace{0.5truecm} &$\epsilon_p$ & \hspace{0.5truecm} &  -3.560 \\
                          & \hspace{0.5truecm} &$\epsilon_z$ & \hspace{0.5truecm} &  -6.886 \\
\\
Intralayer Mo-S & \hspace{0.5truecm} &$V_{pd\sigma}$ & \hspace{0.5truecm} &  3.689 \\
                          & \hspace{0.5truecm} &$V_{pd\pi}$ & \hspace{0.5truecm} &  -1.241 \\
\\
Intralayer Mo-Mo& \hspace{0.5truecm} &$V_{dd\sigma}$ & \hspace{0.5truecm} &  -0.895 \\
                          & \hspace{0.5truecm} &$V_{dd\pi}$ & \hspace{0.5truecm} &  0.252 \\
                          & \hspace{0.5truecm} &$V_{dd\delta}$ & \hspace{0.5truecm} &  0.228 \\
\\
Intralayer S-S & \hspace{0.5truecm} &$V_{pp\sigma}$ & \hspace{0.5truecm} &  1.225 \\
                          & \hspace{0.5truecm} &$V_{pp\pi}$ & \hspace{0.5truecm} &  -0.467 \\
\hline
\hline
\end{tabular}
\caption{Slater-Koster tight-binding parameters for single-layer MoS$_2$.
All  terms
are in units of eV.}
\label{Tab:t-fit}
\end{table}

\subsection{Hamiltonian in strained lattice}
\label{s:strain}

The use of a Slater-Koster tight-binding approach is particularly convenient when lattice deformations, like strain,
are considered.
Neglecting as a first approximation the corrections to the local atomic potentials due to lattice deformation~\cite{JMV13, KG13}
the effect of strain is here driven by the dependence of the tight-binding parameters of the  two-center energy integral elements
which depend on the interatomic distance.
The effect of strain is thus considered in the simplest way by varying the interatomic bond lengths as a result of
the applied strain.
At the linear order, the modified hopping terms in the presence of strain can be written as
\begin{eqnarray}\label{Eq:hopping}
t_{ij,\mu\nu}({\bf r}_{ij})&=&t_{ij,\mu\nu}({\bf r}_{ij}^0)
\left(1-\Lambda_{ij,\mu\nu}
\frac{|{\bf r}_{ij}-{\bf r}_{ij}^0|}{|{\bf r}_{ij}^0|}
\right),
\end{eqnarray}
where $|{\bf r}_{ij}^0|$ is the distance in the equilibrium positions
between two atoms labelled by $(i,\mu)$ and $(j,\nu)$,
$|{\bf r}_{ij}|$ the distance in the presence of strain,
and $\Lambda_{ij,\mu\nu}=-d\ln t_{ij,\mu\nu}(r)/d\ln(r)|_{r=|{\bf r}_{ij}^0|}$
is the dimensionless bond-resolved local electron-phonon coupling.
For practical purposes, we have $|{\bf r}_{ij}^0|=\sqrt{7/12}~a$
for the $M$-$X$ bond, and $|{\bf r}_{ij}^0|=a$
for the in-plane $M$-$M$ and $X$-$X$ bonds.

A microscopic evaluation of the electron-lattice coupling parameters
$\Lambda_{ij,\mu\nu}=-d\ln t_{ij,\mu\nu}(r)/d\ln(r)|_{r=|{\bf r}_{ij}^0|}$
is in principle possible by means of an accurate analysis based on the direct comparison
between  DFT and tight-binding calculations.
Along this line, for instance, the electron-phonon coupling associated with the different interlayer
hopping in multilayer graphene were estimated in Ref. [\onlinecite{CP12}].
This task turns however to be formidable in transition metal dichalcogenides because
of the large number of orbitals/bands and because of the lack of a Fermi surface
that can be used as a reference.
In the absence of any theoretical and
experimental estimation for the electron-phonon coupling,
we use the Wills-Harrison argument~\cite{H99}
namely $t_{ij,\mu\nu}(r) \propto |{\bf r}|^{-(l_\mu+l_\nu+1)}$, where $l_\mu$ is the absolute value of the angular momentum
of the orbital $\mu$, and $l_\nu$ is the absolute value of the angular momentum
of the orbital $\nu$.
Following this approach we assume that
$\Lambda_{ij, X-X}=3$, $\Lambda_{ij, X-M}=4$,
and $\Lambda_{ij, M-M}=5$, for the $X$-$X$ $pp$, for $X$-$M$ $pd$, and
for the $M$-$M$ $dd$ hybridizations, respectively.
The application of strain transforms the vector ${\bf r}_0$, which separates two lattice sites connected with electron hopping,  into ${\bf r}\approx{\bf r}_0+{\bf r}_0\cdot \bm \nabla \bf u$. Note that, in the above transformation, we are considering only the acoustic part of the displacement vector, which has been shown to be a good approximation in the long wavelength region of interest here~\cite{SA02}.

In general, we can write $\bm \nabla\bf u=\bm \varepsilon+\bm \omega$, where $\bm \varepsilon$ and $\bm \omega$ are the strain and rotation tensors, respectively~\cite{KG13}. The strain tensor of a two-dimensional system is given by the symmetric tensor
\begin{eqnarray}
\varepsilon=\begin{pmatrix}\varepsilon_{xx} &&\varepsilon_{xy}\\ \varepsilon_{xy}&&\varepsilon_{yy}\end{pmatrix},
\end{eqnarray}
with components that include both in-plane and out-of-plane displacements
\begin{eqnarray}\label{st}
\varepsilon_{ij}=\frac{1}{2}\left(\frac{\partial u_i}{\partial r_j}+\frac{\partial u_j}{\partial r_i}\right)+\frac{1}{2}\frac{\partial u_z}{\partial r_i}\frac{\partial u_z}{\partial r_j}
\end{eqnarray}
where ${\bf r}=(x,y)$ and ${\bf u}=(u_x,u_y,u_z)$ are the position and displacement vectors, respectively.
The rotation tensor $\bm \omega$ accounts for local rotations in the system. It is an antisymmetric tensor defined by $2\omega_{xy}=-2\omega_{yx}=(\partial u_y/\partial x-\partial u_x/\partial y)$. For a homogeneous strain the rotation tensor is zero and we can assume that ${\bf r}={\bf r}_0\cdot(\bf 1+\bm \varepsilon)$. On the other hand, for an inhomogeneous strain with the local rotation we must use ${\bf r}={\bf r}_0\cdot(\bf 1+\bm\nabla\bf u)$ \cite{KG13}. Explicit expressions for the atomic separation as modified by non-uniform strain are given in Appendix \ref{App:k.p}.

\section{low-energy model of strained TMDs}\label{Sec:k.p}

Hamiltonian (\ref{Eq:H-Real}), which includes explicitly the
hybridization between the metal and the chalcogen atoms,
represents the appropriate starting point for a compelling derivation of an effective
low-energy model in the presence of strain.
For this purpose, we perform a Taylor expansion in momentum and in the
strain fields, followed by a canonical projection onto the two
(conduction and valence) low-energy
bands.
From the technical point of view,
in order to obtain an effective $2 \times 2$
($4 \times 4$ including spin) model Hamiltonian,
we use the L\"{o}wdin partitioning method~\cite{W03}.
Details about the derivation are provided in Appendix \ref{App:k.p}.
Similar to the carbon nanotube and to the graphene
cases~\cite{SA02,SKS05,KN07,GV08}
we first set the momentum coordinates on the relative valley (K-point
of the Brillouin zone), and we derive hence a strain-dependent
Hamiltonian which includes the effect of hopping integrals
modification caused by the deformation. The strain-dependent Hamiltonian
around K-point, up to second order in strain and momentum, can be
written as $H=H_0+H_{\rm so}$, where
\begin{align}\label{Eq:hks}
H_0&=
\frac{\Delta_0+\Delta\sigma_z}{2}+{\cal D}
+t_0 a_0 \left({\bf q}
+\frac{e}{\hbar}\tau{\bf A}_1\right)\cdot{\bm \sigma}_\tau
\nonumber\\&+\frac{\hbar^2}{4m_0}\left(|{\bf q}
+\frac{e}{\hbar}\tau{\bf A}_2|^2\alpha
+|{\bf q}+\frac{e}{\hbar}\tau{\bf A}_3|^2\beta\sigma_z\right),
\nonumber\\
H_{\rm so}&=\left\{\frac{\lambda_0+\lambda\sigma_z}{2}+{\cal \delta\lambda}+a_0^2\left(|{\bf q}+\frac{e}{\hbar}\tau{\bf A}_4|^2\lambda_0'\right.\right.\nonumber\\
&\left.\left.+|{\bf q}+\frac{e}{\hbar}\tau{\bf A}_5|^2\lambda'\sigma_z\right)\right\}\tau s.
\end{align}
Here $e$ is the elementary charge, $m_0$ is the free electron mass,
${\bm  \sigma}_\tau=(\tau\sigma_x,\sigma_y)$
are Pauli matrices in the
$2 \times 2$ ``band'' space, and $s=\pm$ and $\tau=\pm$
are spin and valley indexes, respectively.
Finally
$a_0=a/\sqrt{3}$
and ${\bf q}=(q_x, q_y)$ is the
relative momentum with respect to the K point.
The parameters $\Delta_0$, $\Delta$, $\lambda_0$, $\lambda$,
$\lambda'_0$, $\lambda'$,  $t_0$, $\alpha$ and $\beta$ are
strain-independent
and they can be obtained directly from the Slater-Koster parameters
of the original Hamiltonian (\ref{Eq:H-Real}), and they are given in Table \ref{Tab:t_low} for the case of MoS$_2$. A detailed derivation of the numerical values of all the parameters
of the low-energy model in terms of the original tight-binding
parameters can be found in Appendix \ref{App:k.p}.
\begin{table}[t]
\begin{tabular}{clcl}
\hline
\hline
$\Delta_0=-0.11 eV$ & \hspace{0.5truecm} & $\lambda_0~=69$ ~meV  \\
$\Delta~=~1.82 eV$ & \hspace{0.5truecm} & $\lambda=-81$  meV \\
$\lambda'_0=-17 meV$ & \hspace{0.5truecm} & $\lambda'=-2$ meV \\
\hline
\hline
\end{tabular}
\mbox{}
\vspace{5mm}
\mbox{}
\begin{tabular}{c c c c c c c c}
\hline\hline
&eV&&eV&&meV&&meV\\
\hline
$\alpha^{+}_1$&~~15.99&~~~$\alpha^{-}_1$&~~15.92&~~~$\alpha^{s+}_1$ &~~-61&~~~$\alpha^{s-}_1$&~~-5.7\\
$\alpha^{+}_2$&~~ -3.07&~~~$\alpha^{-}_2$&~~-1.36&~~~$\alpha^{s+}_2$&~~3.2&~~~$\alpha^{s-}_2$&~~ 0.02\\
$\alpha^{+}_3$&~~-0.17&~~~$\alpha^{-}_3$&~~ 0.0&~~~$\alpha^{s+}_3$&~~3.4&~~~$\alpha^{s-}_3$&~~ 0.01\\
\hline
\hline
\end{tabular}
\mbox{}
\vspace{5mm}
\mbox{}
\begin{tabular}{ccccc}
\hline\hline
$\eta_1$ & $\eta_2$ & $\eta_3$ & $\eta_4$ & $\eta_5$ \\
$0.002$ & $-56.551$ & $1.635$ & $1.362$ & $8.180$ \\
\hline
\hline
\end{tabular}
\caption{Microscopical parameters of the spinful two-band low-energy
model. The upper table describes strain-independent
Hamiltonian parameters where $t_0=2.34$ eV, $\alpha=-0.01$ and $\beta=-1.54$; the middle table the Hamiltonian parameters related
to the strain through a scalar potential [Eq. (\ref{Eq:scalar})]; the lower
table the Hamiltonian parameters $\eta_i$ related to the strain-induced coupling to
the pseudo-vector potentials ${\bf A}_i$.}
\label{Tab:t_low}
\end{table}
 It is useful to notice that the mass asymmetry parameter, $\alpha$, and topological term, $\beta$,
are related to general physical properties of the band structure,
like effective mass and energy gap, through the relations
$\alpha=m_0/m_+$ and $\beta=m_0/m_{-}-4m_0v^2/(\Delta-\lambda_{-})$,
where $v=t_0a_0/\hbar$, $m_{\pm}=m_c m_v/(m_v\pm m_c)$, $2\lambda_{\pm}=\lambda_0\pm\lambda$.
In addition, $m_c$ and  $m_v$ are the effective masses of the conduction and
valence band, and $\lambda_+$ and $\lambda_-$ are the spin-orbit
coupling of the conduction and valence bands,
respectively~\cite{RMA13}.

The presence of a finite strain induces in the Hamiltonian (\ref{Eq:hks}) many
different terms. The most straightforward are the {\em diagonal} ones, i.e.
 a scalar potential, which contains a spin independent part, ${\cal D}={\rm diag}[D_{+},D_{-}]$, and a
 spin-dependent contribution, ${\cal \delta\lambda}={\rm diag}[\delta\lambda_{+},\delta\lambda_{-}]$. The
explicit expressions of $D_{\pm}$ and $\delta\lambda_{\pm}$ read:
\begin{eqnarray}\label{def}
D_{\pm}
&=&\alpha^{\pm}_1|{\cal A}|^2+\alpha^{\pm}_2(V+\omega^2_{xy})+\alpha^{\pm}_3 V^2,
\nonumber\\
\delta\lambda_{\pm}
&=&
\alpha^{s\pm}_1|{\cal A}|^2+\alpha^{s\pm}_2(V+\omega^2_{xy})+\alpha^{s\pm}_3 V^2.
\label{Eq:scalar}
\end{eqnarray}
Note that the strain fields appear in Eqs. (\ref{Eq:hks}) and (\ref{Eq:scalar}) only through the
representative variables
${\cal A}=\varepsilon_{xx}-\varepsilon_{yy}-i 2\varepsilon_{xy}$,
$V=\varepsilon_{xx}+\varepsilon_{yy}$,
and $\omega_{xy}=(\partial u_y/\partial x-\partial u_x/\partial y)/2$.
The numerical values of all $\alpha_i$ are also reported in
Table.~\ref{Tab:t_low}. It should be noticed that the quantitative use of  the second order terms in the scalar potential (${\cal D}$ and ${\cal \delta\lambda}$) should be done with some cares. Because, here, we considered the linear approximation to include deformation in the bond lengths. However, these terms would be negligible for small deformation.  

In addition to the above discussed diagonal terms, it is interesting
to underline the appearance  in (\ref{Eq:hks}) of  {\em five}
different fictitious gauge fields defined as
${\bf A}_i=\eta_i{\bf A}$, where
$A_x=(\hbar/e a_0){\rm Re}[{\cal A}]$ and $A_y=(\hbar/e a_0){\rm Im}[{\cal A}]$.
The coupling constants $\eta_i$ are evaluated from the
values of the initial Slater-Koster parameters, and their specific value
for the case of single-layer MoS$_2$ are reported in Table \ref{Tab:t_low}.
Note that, due to the small value of $\eta_1$,
the off-diagonal pseudo vector potential  (${\bf A}_1$) results to be
very weak as compared to the diagonal ones. The opposite happens for
the well known cases of mono- and bilayer graphene, for which the
off-diagonal terms are the dominant components of the strain dependent
Hamiltonian~\cite{SA02,VKG10}. The weakness of ${\bf A}_1$ in MoS$_2$
might stem from the large energy gap as compared with graphene,
which is a semi-metal with no gap.

\section{Strained TMDs as a multi-pseudo-vector field system}\label{Sec:Pseudo}

The dependence of the electronic/transport/optical properties of TMDs
triggers the biggest interest towards realistic applications for
strain-engineering in these materials and hence many theoretical and experimental setups have been proposed.
Most of them employ the dependence of the {\em magnitude} of the
optical or transport gap. On the conceptual basis, such proposals are
thus related to the strain modulation of the {\em scalar} potentials
$D_\pm$, $\delta\lambda_\pm$. Interesting enough, there is, on the other hand, 
off-diagonal terms that can be described in terms of the pseudo-vector potential.
The concept of pseudo gauge fields, for instance,
has been widely discussed in the context of strained graphene~\cite{SA02,SKS05,KN07,VKG10}
and it provides the possibility to induce extremely large effective
pseudomagnetic fields~\cite{LC10}.
Such pseudomagnetic fields are thus reflected in the onset of flat
bands (Landau levels)
in the energy spectrum of deformed systems,
as observed experimentally in strained graphene
samples~\cite{LC10,Gomes12,Lu12}.
A similar framework has been discussed in TMDs by Cazalilla {\it et al.}~\cite{COG14}.
However, the Hamiltonian in Eq. (\ref{Eq:hks}), appropriate for realistic
modelling of monolayer TMDs, shows profound differences in regards
to this simplistic picture since at least
{\em three} pseudo vector potentials (${\bf A}_1$, ${\bf A}_2$, and
${\bf A}_3$) are induced by strain, even in the simplest spinless case.
It should be stressed that,
due to this multi-pseudo-vector field structure,
these fields {\em cannot} be referred as {\it gauge fields}, since
the contemporary presence of {\em three} fields  cannot
be described by
a simple phase shift in the wave-function
after doing a transformation like  ${\bf A}\rightarrow{\bf A}+\nabla \Lambda$.
In other words, the effect of  these pseudo vector fields can not be eliminated by
the counter-acting presence of a {\em real} magnetic field.

The complexity of these multi pseudovector field structures
gives rise to qualitatively new physics, which is not present in
graphene-like systems and in the previous analyses for TMDs based on
only one vector potential.
The rich phenomenology of this structure will be thus the object of investigation
of the present Section.

An interesting feature in graphene under strain is the possibility
of tailoring the pseudo vector potential in such a way to mimic the effects of
an effective magnetic field. Under these conditions, (pseudo-) Landau levels
are expected to appear, reflected in flat bands in the electronic band
structure. This scenario has been theoretically predicted~\cite{GKG10}  and experimentally
verified in graphene~\cite{LC10} and similar predictions have been prompted out
in TMDs~\cite{COG14}.
Things are actually more
complex in a realistic modelization of the TMDs, due to the presence of
many pseudo-vector potentials as we will discuss later.

In order to focus on the possible occurrence of pseudo Landau levels
(PLLs), we neglect in the following the role of the scalar potentials,
and we consider only the leading term of the spin-orbit coupling
in the absence of strain. Only three pseudo vector potentials will appear, ${\bf A}_j$  with $j=1,2,3$.
The first standard step to address this issue is to introduce
the total {\it canonical} momentum fields
${\bm\pi}_j=(\hbar{\bf q}+e{\bf A}_j)=(\hbar{\bf q}+e\eta_j{\bf A})$,
where ${\bm \pi}_j={\pi}^x_j+i{\pi}^y_j$.
The fundamental thing to be underlined here is that the fields ${\bm \pi}_j$ are {\rm not}
orthogonal, but fulfill the following relations:
\begin{align}
\label{pipj}
[\pi_i, \pi_j] &=
-ie\hbar(\eta_j-\eta_i)(\partial_x+i\partial_y)(A_x+iA_y)
\nonumber \\
[\pi_i,\pi^\dagger_j ] &=
-ie\hbar(\eta_j-\eta_i){ \nabla }\cdot{\bf A}-e\hbar(\eta_j+\eta_i)({ \nabla}\times {\bf A})_z.\nonumber\\
\end{align}

It is interesting to notice that  when all  vector potentials have the same
couplings $\eta_i=\eta$, then there is just one gauge field, which leads to the standard algebra for the canonical momentum associated with a real magnetic field. Therefore, this kind of solution corresponds to the {\it real} Landau levels of the system in the presence of a {\it true} magnetic field.

However, the commutation relations in (\ref{pipj}) for the more general case of strained TMDs, imply that such operators do {\em  not} commute
and obey a more complex algebra.
Finding an analytical solution for the PLLs results thus a formidable task due to the
non-orthogonality of the theory~\cite{EL91} which is one of the
main consequences of the presence of multi-pseudo-vector fields. Nevertheless, it is instructive to consider the symmetric gauge, ${\bf A}=\frac{B}{2}(-y, x)$, which is the one associated with the experimentally relevant case of trigonal deformation of the lattice~\cite{GKG10,LC10}.
In this case one can show that
\begin{align}
[\hat{a}_i,\hat{a}^\dagger_j]=S_{ij}~~~,~~~[\hat{a}_i,\hat{a}_j]=0~~~,~~~[\hat{a}^\dagger_i,\hat{a}^\dagger_j]=0
\end{align}
where $S_{ij}=(\eta_i+\eta_j)/2\sqrt{|\eta_i\eta_j|}$ and we have introduced the creation operators
$\hat{a}^{\dagger}_i=\frac{l_{B}}{\sqrt{2|\eta_i|}\hbar}{ \pi}_i$, where $l_{B}=\sqrt{\hbar/e|B|}$ is the magnetic length in terms of the
pseudo-magnetic field $B=|\nabla\times{\bf A}|$. The overlap matrix of this case, $S_{ij}\neq \delta_{ij}$, implies that the bosonic operators $a_i$
are in general non-orthogonal, and $[\hat{n}_{i},\hat{n}_{j}]\neq 0$
for $i\neq j$. In order to give a solution for this problem, one needs to redefine Fock space in a non-orthogonal basis~\cite{EL91}. In Sec. \ref{Sec:SingleBand} we will provide with a perturbative analytical solution using a single band model.

\subsection{Energy spectrum in the Landau gauge: arc-shape deformation}

\begin{figure}
\centering
\includegraphics[width=0.8\linewidth]{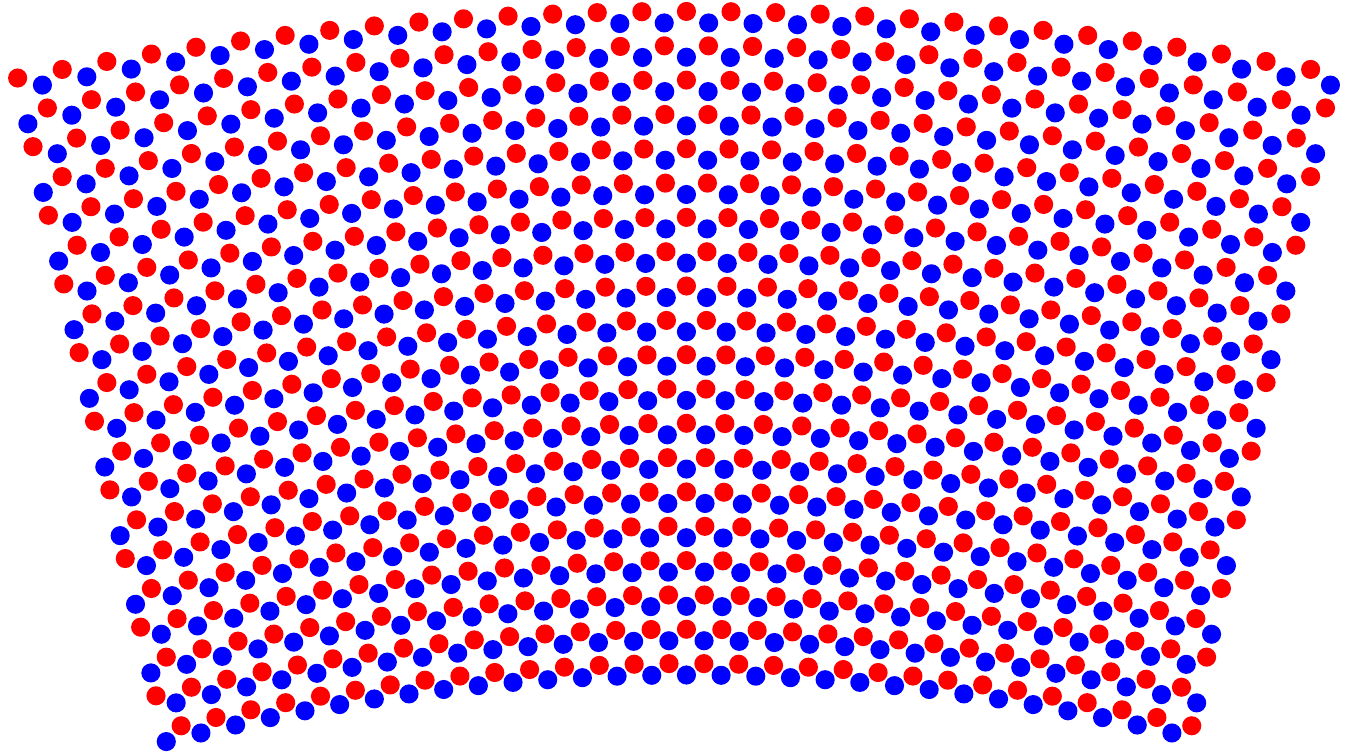}
\caption{(Color online) Top view of an arc-shaped $MX_2$ with $R=4L_y$ in which blue and red lattice points indicates $M$ and $X$ atoms, respectively.}
\label{Fig:arc-shape}
\end{figure}

Besides this specific case, however, the complex multi-vector potential
structure of the Hamiltonian does not allow for an analytical determination of
the energy  bands and of the pseudo Landau levels.
We have thus solved the problem numerically.
In order to reveal the relevant physics associated with
the pseudo-vector potentials and with possible pseudo-Landau levels,
we choose an inhomogeneous strain profile corresponding to a constant pseudo magnetic
field for each vector potential. An arc-shape deformation (sketched in Fig. \ref{Fig:arc-shape}),
which corresponds to a displacement profile $(u_x,u_y)=(xy/R,-x^2/2R)$ ($R$ being here
the arc radius), is known to be one of the simplest and efficient candidates~\cite{LG10,RA13b}.
Within this context the resulting gauge field in the Landau gauge reads
 ${\bf A}=\frac{\hbar}{ea_0}(y/R,0)$, corresponding to a
three different constant pseudo magnetic field
$B_i =\eta_i\hbar/e a_0 R$, associated to the pseudo vector potentials ${\bf A}_i=\eta_i{\bf A}$.
Neglecting the weak contribution of the rotation tensor, Hamiltonian (\ref{Eq:hks}) can be written in first-quantization as
\begin{widetext}
\begin{eqnarray}
\label{hpll2}
H=\begin{pmatrix}V_{+}(y)-\frac{\hbar^2}{4m_0}(\alpha+\beta)\partial_y^2&&t_0a_0(q_x+\eta_1\frac{y}{a_0R})-t_0a_0\partial_y\\
\\t_0a_0(q_x+\eta_1\frac{y}{a_0R})+
t_0a_0\partial_y&&V_{-}(y)-\frac{\hbar^2}{4m_0}(\alpha-\beta)\partial_y^2\end{pmatrix},
\end{eqnarray}
where
\begin{eqnarray}
V_{\pm}(y)=\frac{(\Delta_0+\lambda_0 s)\pm(\Delta+\lambda s)}{2}+D_{\pm}(y)+\delta\lambda_{\pm}(y)s
+\frac{\hbar^2}{4m_0}\alpha\left(q_x+\eta_2 \frac{y}{a_0R}\right)^2\pm \frac{\hbar^2}{4m_0}\beta\left(q_x+\eta_3 \frac{y}{a_0R}\right)^2.
\end{eqnarray}
\end{widetext}
The solution of the above Hamiltonian leads to a set of coupled differential equations that we solve numerically for hard-wall boundary conditions. Details can be found in Appendix \ref{numerical-fem}.
\begin{figure}
\centering
\includegraphics[width=0.7\linewidth]{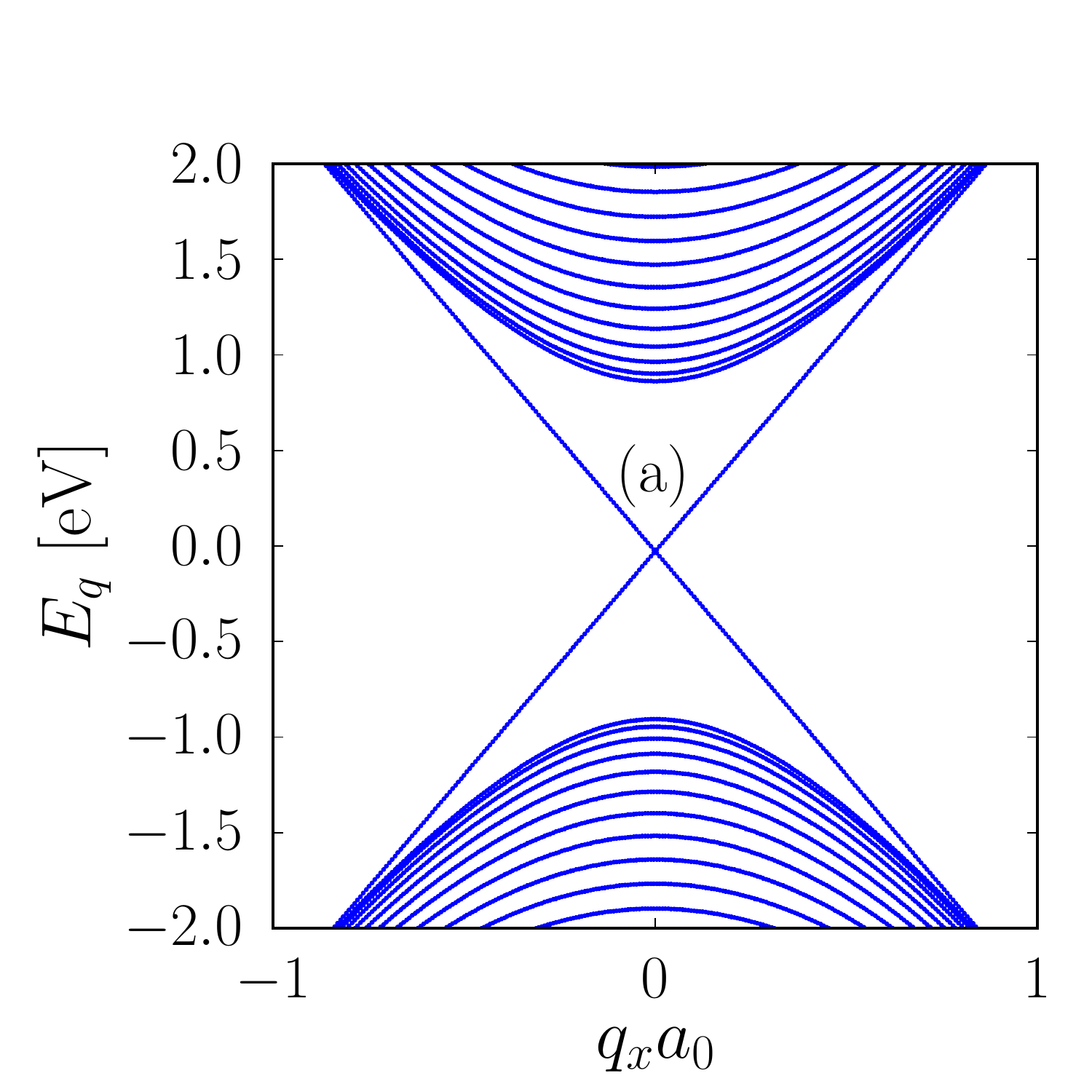}
\includegraphics[width=0.7\linewidth]{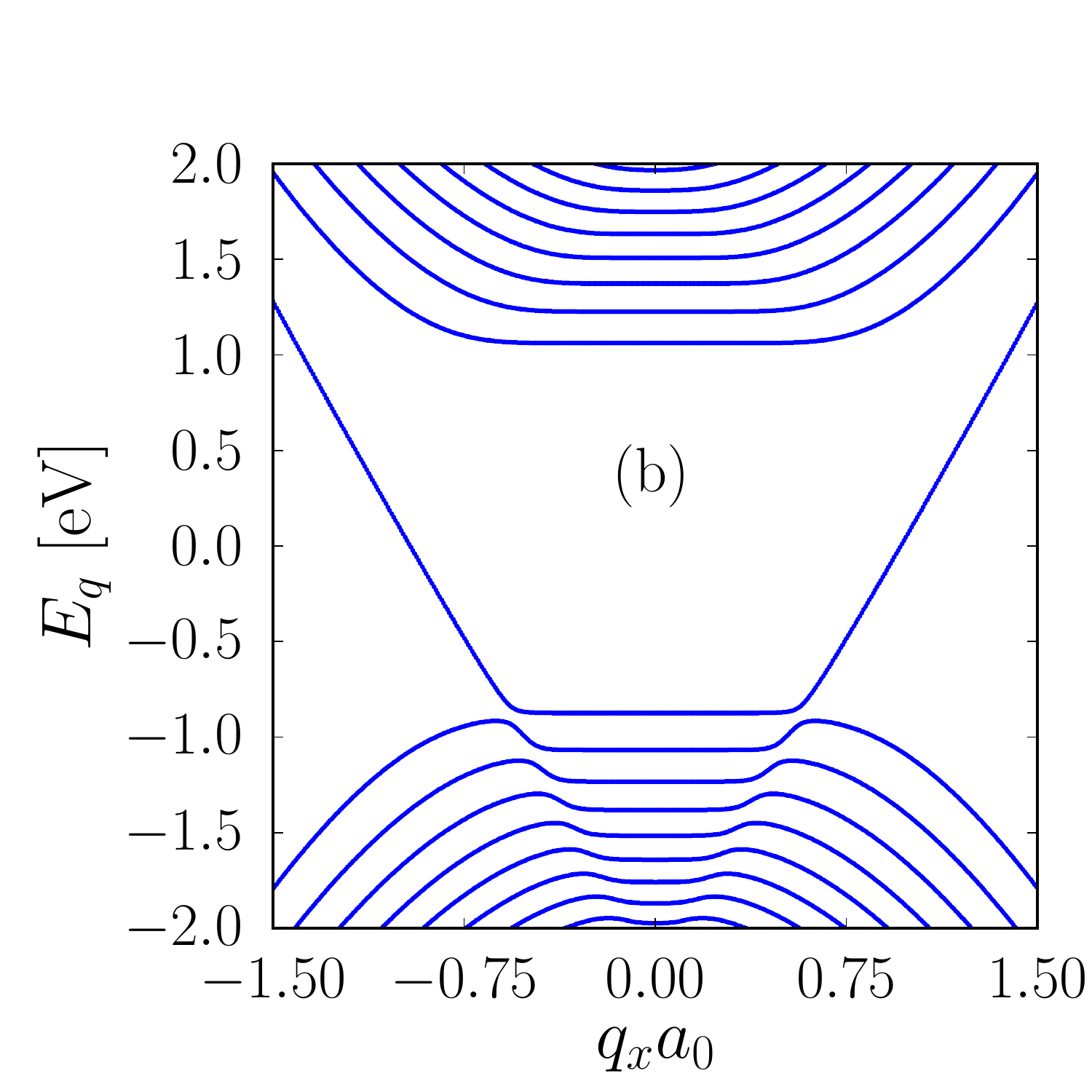}
\caption{(Color online) (a) Energy dispersion of unstrained nanoribbon
  case with $L_y=50a_0$.
(b) Energy dispersion a strained nanoribbon for
$L_y=50a_0,~\eta_i=\eta=1$ and $R=0.5L_y$. Noticeably, the first Landau
level in the valence band evolves in an electron-like edge
mode.}
\label{fig2}
\end{figure}

\begin{figure}
\centering
\includegraphics[width=0.7\linewidth]{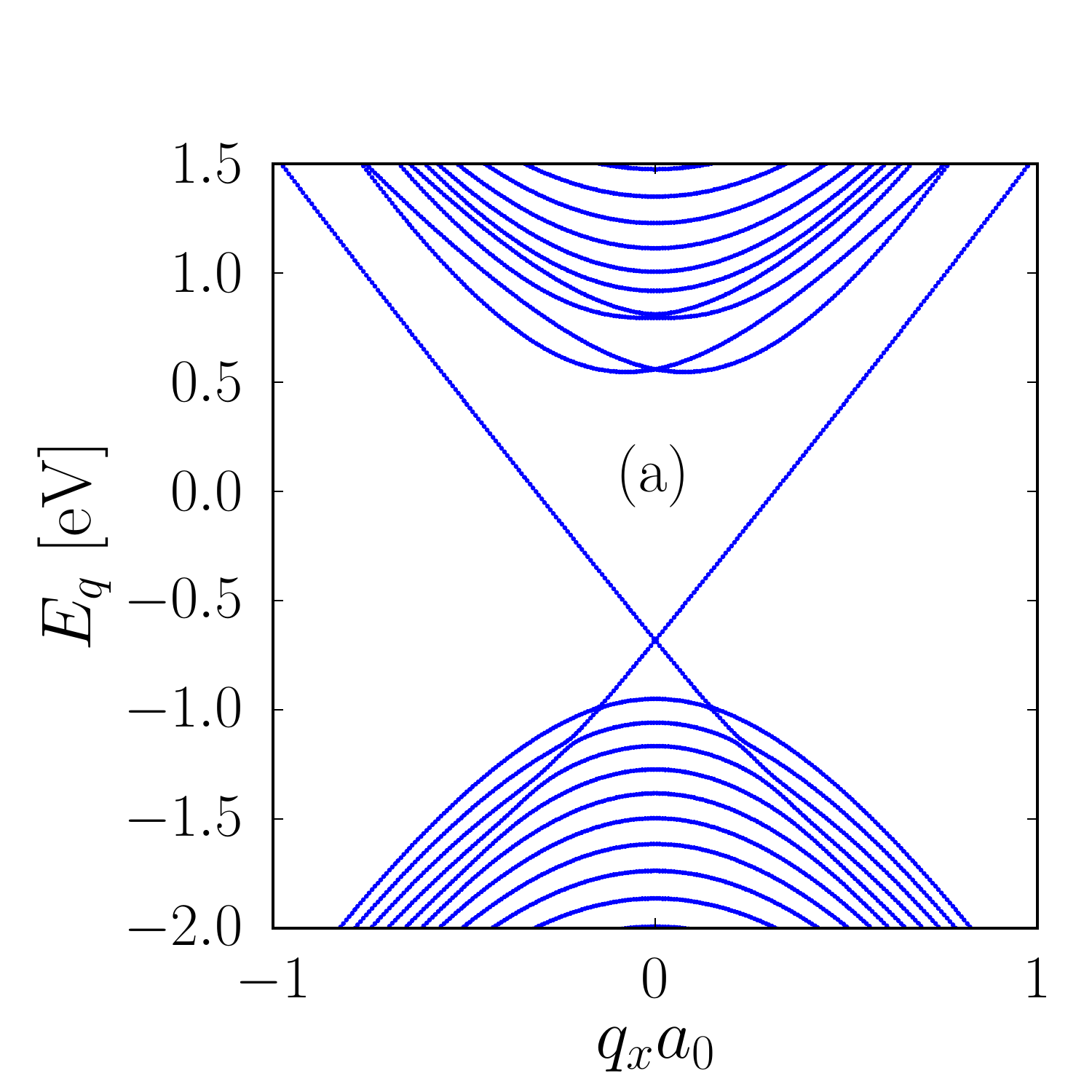}
\includegraphics[width=0.7\linewidth]{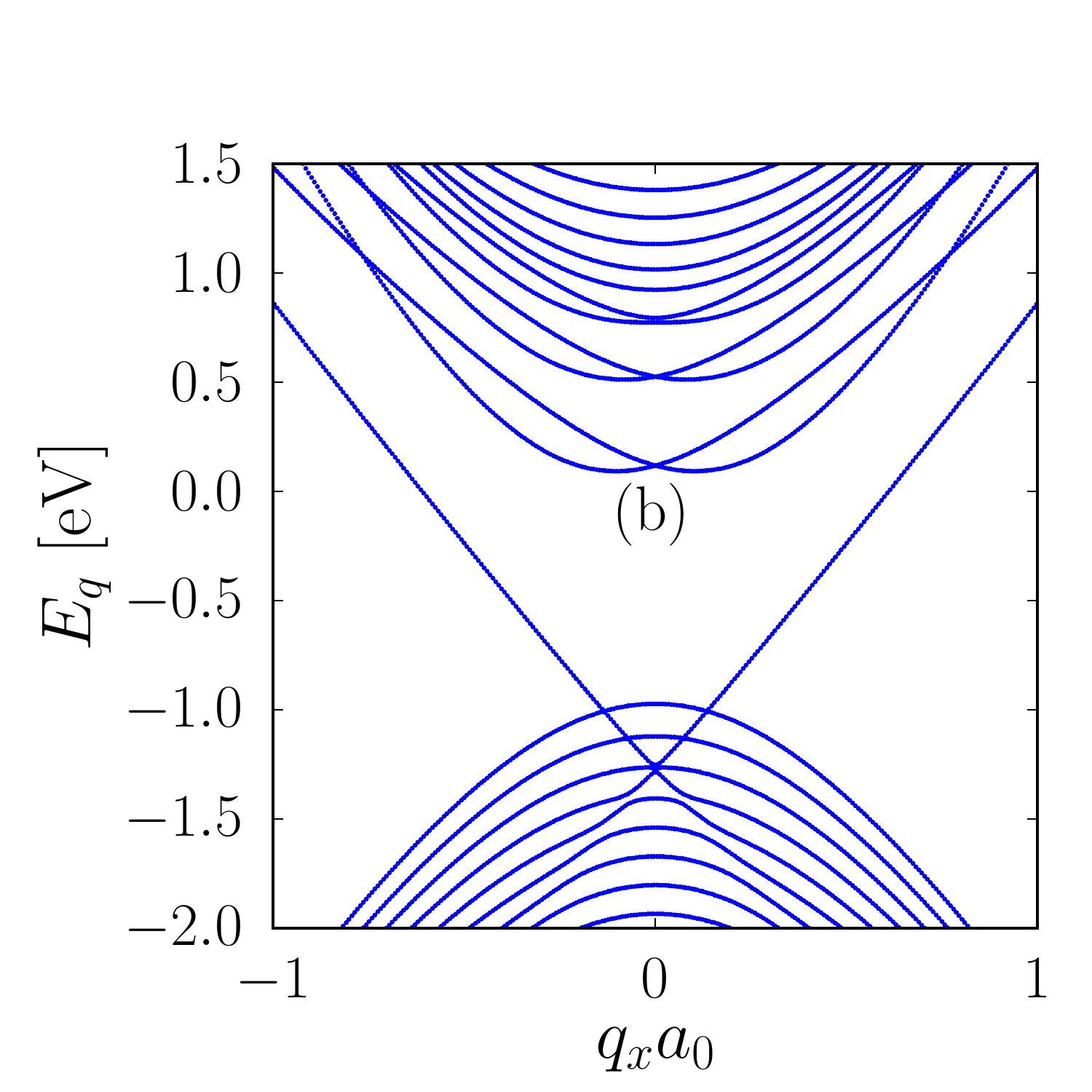}
\caption{(Color online)  Energy dispersion of arc-shaped case with
  scalar potential and rotation tensor neglected. The dispersion relation
show equidistant parabolic valence subbands, and
a set of conduction subbands that present band crossing
for the low-energy states and parabolic dispersion
above some specific energy. The size of the energy gap and level spacing between
the subbands is strongly dependent on the strain strength. We set (a)$ L_y=50a_0,R=2.5L_y$ and (b) $L_y=50a_0,R=1.8L_y$.}
\label{fig3}
\end{figure}
The resulting dispersion relations under different strain conditions are shown in Fig. \ref{fig2} and
 (\ref{fig3}) where, in order to analyze the possible presence of pseudo-Landau levels,
the scalar potentials ($ D_{\pm}$ and $\delta\lambda_{\pm}$) have been neglected.
In comparison, we show the band dispersion of the unstrained case in Fig. (\ref{fig2})a.
Apart from the parabolicity of the valence and conduction bands, one can observe a crossing of the edge modes. This is expected due to the nonzero Chern number associated to each flavour (spin or valley) in our model
(i.e. $2{\cal C}={\rm sign}(\Delta)-{\rm sign}(\beta)$)~\cite{RA14}. 
In other words, as long as the valley index is a good quantum number (for instance in zigzag termination and hard-well boundary cases), and the valley-Chern number is {\it integer valued}, then metallic edge modes are expected to exist in the gap.
Although this problem is beyond the scope of the present paper, the nature of these edge states will be discussed in Sec.~\ref{Sec:Scalar} using the numerical results of TB model. Here we mention that the existence of metallic edge modes in this kind of systems depend on the edge potential and atomic termination.\cite{LM12,YN09}

Fig. \ref{fig2}b shows the band dispersions in the previously
discussed toy model where the coupling of all the three strain induced fields
is the same ($\eta_i=\eta$), which correspond to the case of a real magnetic field applied to the sample.
Remarkably, the first Landau level in the valence band evolves in an
electron-like edge mode.
This feature is consistent with the results of the full six-band
tight-binding model~\cite{RA15}.

The presence of flat bands (pseudo-Landau levels) can be speculated
from these results.
However, as we are going to show, the actual relevance of pseudo-Landau
levels appear doubtful in the more realistic case of arc-shaped tension.
In Fig. \ref{fig3} we set the parameters $\eta_i$ with the numerical
values listed in Table \ref{Tab:t_low}, which correspond to the realistic
full tight-binding dispersion.
Panels (a) and (b) show the band dispersion of a given spin flavour for two different magnitudes of strain,
parametrized in terms of two values of the arc radius $R$.
The resulting dispersion show almost equidistant parabolic valence
subbands, and a set of conduction subbands that present band crossing for the low-energy states and parabolic dispersion above some specific energy.
It should be noticed also that the size of the energy gap and level
spacing between the subbands (in both valence and conduction sectors)
is strongly dependent on the strain strength.
In the following section we will show how these peculiar features, for
both valence and conduction bands, can be understood in terms of a
harmonic oscillator and inverted harmonic oscillator physics, respectively.

In summary, comparing these results in strained MoS$_2$ (as a representative case of single layer TMD), with  those of strained graphene~\cite{VKG10} reveals three main differences between these two systems: $i$) No obvious similarity between pseudo vector potentials and real magnetic field, as it happens in strained graphene. $ii$) Strong particle-hole asymmetry in the strained MoS$_2$ energy dispersion, as compared to the symmetric spectrum in strained graphene. $iii$) Absence of flat PLL in MoS$_2$ for Landau's gauge (i.e. arc-shape) deformation in contrast with the appearance of flat bands in the arc-shaped graphene~\cite{LG10,RA13b}.

\section{Single-band (effective mass) model for strained TMDs}\label{Sec:SingleBand}

The two-band Hamiltonian (\ref{Eq:hks}) can be perturbatively decomposed into two individual one-band Hamiltonians for
the conduction and the valence bands at small $q$ and strain. Those single-band Hamiltonian which is valid around the $K$ point of the BZ, have the following analytical expressions (see Appendix \ref{derivation-single-band} for a detailed derivation)
\begin{align}\label{Eq:h-single}
H_{\pm}&=E_{\pm}+D_{\pm}+\frac{\hbar^2}{4m_0}\{\alpha|{\bf q}+\frac{e}{\hbar}{\bf A}_2|^2\pm\beta|{\bf q}+\frac{e}{\hbar}{\bf A}_3|^2\nonumber\\&\pm\gamma|{\bf q}+\frac{e}{\hbar}{\bf A}_1|^2\}
\end{align}
where $E_{\pm}=\frac{1\mp1}{2}s\lambda_{-}+\frac{\Delta_0\pm\Delta}{2}\pm\eta_1\frac{(t_0 a_0)^2}{(\Delta-s\lambda_{-})l_B^2}$, $\gamma=4m_0 v^2/(\Delta-s\lambda_{-})$, $s$ is the spin index and $+ (-)$ indicates conduction (valence) band, respectively.
Having neglected $\alpha$ and $\beta$ terms, we could get the model Hamiltonian used in Ref.~[\onlinecite{COG14}] to study a pseudomagnetic field in a monolayer MoS$_2$, which reveals (in the absence of SOC) symmetric PLL in the conduction and valence bands. Notice that the single band model (\ref{Eq:h-single}), considers the effective mass asymmetry ($\alpha$) and momentum dependent mass term ($\beta$), leading to different model solutions for the two cases.

 \par
The Hamiltonian from (\ref{Eq:h-single}) can be easily deduced for the Landau's gauge (arc-shape) deformation and by neglecting scalar potential contribution $D_{\pm}$ we have 
\begin{align}\label{eq:arc-perturb}
H_{\pm}=E_{\pm}+\frac{\hbar^2}{4m_0}[(\alpha\pm\beta\pm\gamma)q^2_y +
w^{\pm}_1(y-w^{\pm}_2 q_x)^2+w^{\pm}_3 q^2_x]
\end{align}
where
\begin{align}
w^\pm_1&=\frac{1}{a^2_0R^2}[\alpha\eta^2_2\pm\beta\eta^2_3\pm\gamma\eta^2_1]\nonumber\\
w^\pm_2&=-a_0R\frac{\alpha\eta_2\pm\beta\eta_3\pm\gamma\eta_1}{\alpha\eta^2_2\pm\beta\eta^2_3\pm\gamma\eta^2_1}\nonumber\\
w^\pm_3&=\alpha\pm\beta\pm\gamma-\frac{(\alpha\eta_2\pm\beta\eta_3\pm\gamma\eta_1)^2}{\alpha\eta^2_2\pm\beta\eta^2_3\pm\gamma\eta^2_1}
\end{align}
\par
In principle, the low-energy Hamiltonian reveals two possible scenarios, for the case of inhomogeneous arc-shaped strain which provides a Landau's gauge for the pseudo vector potentials, depending on the sign of $(\alpha\pm\beta\pm\gamma)w_1^{\pm}$ in the single-band models. One is a harmonic oscillator (HO) physics where $(\alpha\pm\beta\pm\gamma)w_1^{\pm}>0$ and the other is a double quantum well (DQW) physics where $(\alpha\pm\beta\pm\gamma)w_1^{\pm}<0$. In the case of MoS$_2$ which is addressing here, after plugging the numerical value of the model parameters in (\ref{eq:arc-perturb}), we obtain
$(w^{+}_1,w^{-}_1)\simeq(-32.2,-24.0)\times(a_0R)^{-2}$,
$(w^{+}_2,w^{-}_2)\simeq(0.06,0.12)\times(a_0R)$, $(w^{+}_3,w^{-}_3)\simeq(4.05,-3.57)$, $\alpha+\beta+\gamma=3.92$ and $\alpha-\beta-\gamma=-3.94$ for the up component of spin index. Therefore, the model for MoS$_2$ leads to a DQW and HO for the conduction and valence bands, respectively.

We would like to emphasize that the single band Hamiltonian (\ref{eq:arc-perturb}) provides with a general model that can be applied to other families of strained semiconductor 2D crystals. For the sake of completeness, we describe the generalities of the model. First of all, we point out that $\alpha+\beta+\gamma>0$ and $\alpha-\beta-\gamma<0$, since these signs originate from the positive and negative masses of electrons and holes, respectively. These relations imply that $0<|\alpha|<\beta+\gamma$ which imposes some restriction over the two-band model parameters just based on the sign of effective masses.

The sign of $w_1^{\pm}$, however, might change depending on the microscopic properties of the considered system. Furthermore, if the effective masses are the same for both bands (i.e. $\alpha=0$) then $w_1^{-}=-w_1^{+}$. In this case, there are two possibilities, namely $w_1^{+}>0$ ($w_1^{+}<0$) which leads to HO (DQW) solutions for both the valence and conduction bands. Therefore, an asymmetry in the effective masses is necessary to have two different physics (understood as HO or DQW energy spectra) in the electron and hole bands.

Finally, we briefly discuss the case of trigonal deformation $ (u_x, u_y) = \frac{u_0}{2} \left(xy,\frac{x^2-y^2}{2}\right)$, where $u_0$ have the units of inverse length and quantifies the strength of the applied strain. This strain profile has been widely discussed in the context of graphene~\cite{GKG10,LC10} and recently it has been considered in TMDs~\cite{COG14}. Such a deformation is properly described by the symmetric gauge, i.e. ${\bf A}=\frac{\hbar u_0}{ea_0}(y,-x)$, which leads to the single band Hamiltonian
 \begin{align}\label{Eq:HSymm}
H_{\pm}&=E_{\pm}+\frac{\hbar^2}{4m_0}[(\alpha\pm\beta\pm\gamma)(q^2_x+q^2_y)
\nonumber\\&
+ z^{\pm}_1 (x^2+y^2) -2 z^{\pm}_2 (x q_y-y q_x)]
 \end{align}
where $z^{\pm}_1= \frac{u_0^2}{a^2_0}[\alpha\eta^2_2\pm\beta\eta^2_3\pm\gamma\eta^2_1]$ and $z^{\pm}_2= u_0 (\alpha \eta_2 \pm\beta \eta_3\pm\gamma \eta_1)$. Notice that since $z_1^{\pm}<0$ in the case of MoS$_2$, the energy spectrum obtained from (\ref{Eq:HSymm}) corresponds to the HO in the valence band and DQW in the conduction band, in a similar manner than in the case of the Landau gauge discussed before. It is, however, important to notice that the lack of the quadratic modulation, i.e. $w^{\pm}_3 q^2_x$, of the HO band in Eq.~(\ref{Eq:HSymm}) results in the appearance of the flat PLL in the valence band, as it has been discussed by Cazalilla {\it et al.}~\cite{COG14}. This has to be compared to the case of the arc-shaped deformation, for which such a quadratic term appears in (\ref{eq:arc-perturb}) through $w_3^{\pm}$, and leads to a set of equidistant parabolic bands in the spectrum.

\subsection{Harmonic Oscillator solution in the valence band}

Using the single-band model for an arc-shaped deformation given by Eq.~(\ref{eq:arc-perturb}), it is possible to find an approximated analytical expression for the energy dispersion shown in Fig. \ref{fig3}. By using the basic physics of harmonic oscillator, it is straightforward to obtain the characteristic equidistant parabolic bands of this system, as given by
\begin{align}\label{Eq:LLv}
E_v(n,q_x)&=E_{-}-\frac{\hbar^2}{2m_0}\sqrt{|w^{-}_1(\alpha-\beta-\gamma)|}\left(n+\frac{1}{2}\right)
\nonumber\\&
+\frac{\hbar^2}{4m_0}w^{-}_3q^2_x
\end{align}
where $n=0,1,2,\dots$. Such a spectrum is consistent with the numerical result shown in Fig. \ref{fig3}. A simple estimation based on Eq. (\ref{Eq:LLv}) suggests a separation between subbands of the order of $\Delta E_v\approx 11.125 \frac{a_0}{R}$~eV.  Since the maximum strain in the arc-shaped system is about $\varepsilon_{max}\approx L_y/2R$, therefore
$\frac{\Delta E_v}{k_{\rm B} T}\approx 860 \times \frac{300}{T}\times \frac{a_0 \varepsilon_{max}}{L_y} $. For a low temperature $T=5$K and given system size $L_y\approx 50$ nm and strain strength about $\varepsilon_{max}=0.1$, we have $\Delta E_v \approx 20~ k_{\rm B} T$ which implies sufficiently spaced levels as to be observed via STM spectroscopy.

\subsection{Double quantum well solution in the conduction band}

\begin{figure}
\centering
\includegraphics[width=0.7\linewidth]{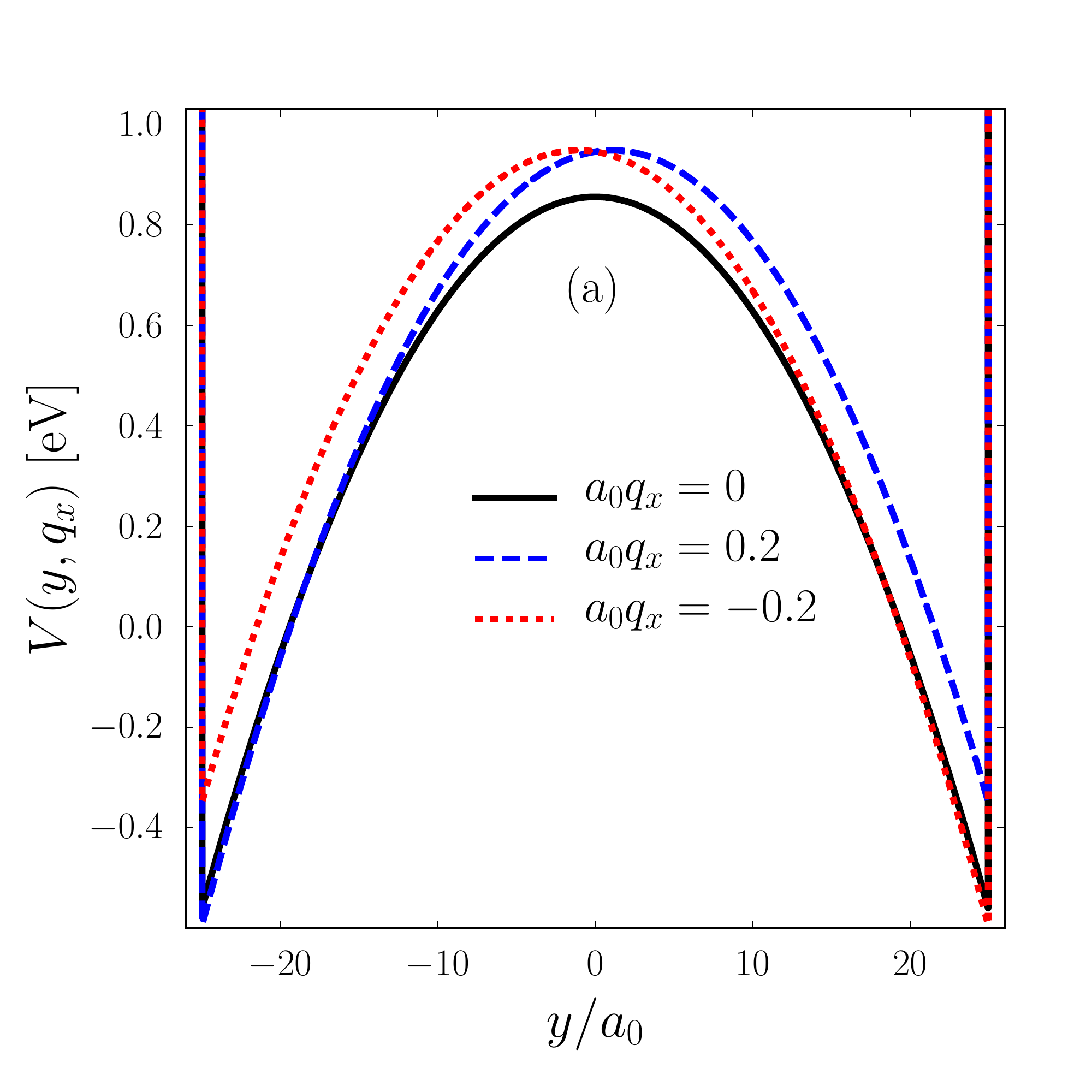}
\includegraphics[width=0.7\linewidth]{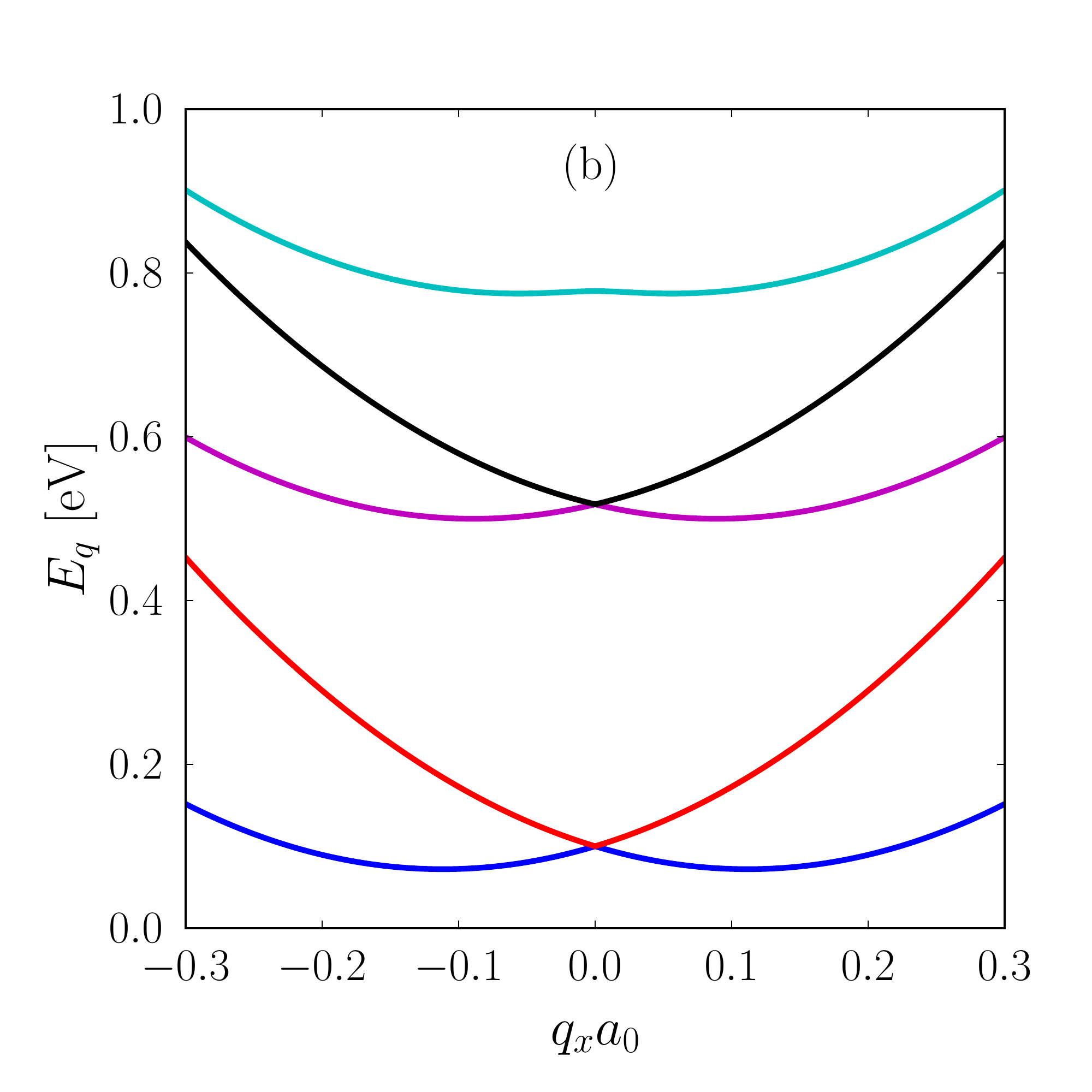}
\caption{(Color online) (a) Potential profile in the conduction band. (b) Five low-energy bands around the K-point in the conduction band, as calculated from the single-band model (\ref{Eq:HDQW}) for the parameters $L_y=50a_0$ and $R=1.8L_y$.}
\label{fig4}
\end{figure}

In order to understand the energy dispersion of the conduction band, we first include two hard walls at $y=\pm L_y/2$ which lead to the following Hamiltonian
\begin{align}\label{Eq:HDQW}
H_{+}=\frac{\hbar^2}{4m_0}(\alpha+\beta+\gamma)q^2_y+ V(y,q_x)
\end{align}
where
\begin{align}
V(y,q_x)&=E_{+}+\frac{\hbar^2}{4m_0}w_3^{+}q^2_x+\frac{\hbar^2}{4m_0}w^{+}_1(y-w^{+}_2q_x)^2
\nonumber\\&
+V_0\left\{1+\Theta\left(y-\frac{L_y}{2}\right)-\Theta\left(y+\frac{L_y}{2}\right)\right\}\nonumber\\
\end{align}
where $\Theta(x)$ is the step function, and $V_0\gg1$ stands for the hard wall potential.
Such hard-well boundary condition can be realized by using external gates. 
Moreover, we expect that this boundary condition is justified 
for ribbons whose termination does not mix the valley degree of freedom, like the zigzag ribbon case. However, as it has been recently shown in Ref. \onlinecite{PB15}, the boundary condition in the continuum model of TMDs is not as simple as graphene case. This is so because the basis spinors in graphene are associated to the sublattice degree of freedom, while in monolayer TMDs they account for the conduction and valence band basis. 
The potential profile is shown in Fig. \ref{fig4}(a) for different values of $q_x$. Notice that the potential is a symmetric (an asymmetric) double quantum well potential for $q_x=0~(q_x\neq0)$  with a barrier in the middle of the sample and two wells located at the edges.
Therefore, the appearance of two wells close to the boundaries indicates the formation of a double quantum well (DQW) in the conduction band.
According to Fig. \ref{fig3}, the energy dispersion becomes parabolic for energies higher than a certain critical value at $q_x=0$. This feature obviously depends on the height of the parabolic barrier $V(y,q_x)$. In fact, for energies higher than the barrier height, which is $\sim18.4~{\rm eV}\times(L_y/2R)^2$, carrier motion does not be much affected by the existence of the barrier.

At finite momentum $q_x$ the energy difference between two minima at $y=\pm L_y/2$ is given by
\begin{align}
\delta E&=V(\frac{L_y}{2},q_x)-V(-\frac{L_y}{2},q_x)=-\frac{\hbar^2 w^{+}_1w^{+}_2 L_y}{2m_0}\times q_x
\end{align}
which mimics an asymmetric DQW at any finite $q_x$ for which the two wells are no longer identical.

If we neglect the hard-wall boundary condition, the Hamiltonian will be exactly solvable for the conduction band (\ref{Eq:HDQW}) which corresponds to the inverted harmonic oscillator equation~\cite{YC06,BF13}. By performing  elementary quantum mechanical approaches, we find the following eigenvalue relation for the wave function and corresponding energy eigenvalues in the conduction band
\begin{align}
\left[\frac{d^2}{dz^2}+\frac{1}{4}z^2\right]\phi(z,\tilde\varepsilon)=\tilde\varepsilon\phi(z,\tilde\varepsilon)
\end{align}
where
\begin{align}\label{eq:z-vare}
&z=\sqrt{\omega}(y-w^{+}_2q_x)\nonumber\\
&\tilde\varepsilon=\frac{-1}{\omega(\alpha+\beta+\gamma)}\left[\frac{4m_0(E-E_{+})}{\hbar^2}-w^{+}_3 q_x^2\right]
\end{align}
and $\omega=2\sqrt{-w^{+}_1/(\alpha+\beta+\gamma)}$. This differential equation is one of the standard differential equations for the parabolic cylindrical functions~\cite{AS65} whose two independent solutions can be written in terms of the confluent hypergeometric function, $M(a,b,c)$
\begin{align}
\phi_{\rm even}(z,\tilde\varepsilon)&=e^{-i\frac{z^2}{4}} M\left(-i\frac{\tilde\varepsilon}{2}+\frac{1}{4},\frac{1}{2},i\frac{z^2}{2}\right)\nonumber \\
\phi_{\rm odd}(z,\tilde\varepsilon)&=ze^{-i \frac{z^2-\pi}{4}} M\left(-i\frac{\tilde\varepsilon}{2}+\frac{3}{4},\frac{3}{2},i\frac{z^2}{2}\right)
\end{align}
which are even and odd functions with respect to the parity operator. The general solution of this problem is a superposition of even and odd eigenfunctions, $\phi=c_1 \phi_{\rm even}+c_2 \phi_{\rm odd}$, where $c_{1,2}$ are unknown constants. To find the corresponding eigenvalues, we must apply the hard-wall boundary condition which implies that the wave function must be zero at $y=\pm L_y/2$. In this regard, it is easy to find the following relation for the eigenvalue problem
\begin{align}\label{eq:boundary}
\frac{\phi_{\rm even}(\sqrt{\omega}(\frac{L_y}{2}-w^{+}_2 q_x),\tilde\varepsilon)}{\phi_{\rm odd}(\sqrt{\omega}(\frac{L_y}{2}-w^{+}_2 q_x),\tilde\varepsilon)}
=-\frac{\phi_{\rm even}(\sqrt{\omega}(\frac{L_y}{2}+w^{+}_2 q_x),\tilde\varepsilon)}{\phi_{\rm odd}(\sqrt{\omega}(\frac{L_y}{2}+w^{+}_2 q_x),\tilde\varepsilon)}
\end{align}
At $q_x=0$ this eigenvalue problem reduces to searching for zeros of the confluent hypergeometric function. These solutions can be labeled as $\tilde\varepsilon=\tilde\varepsilon_n$. If the even wave function satisfies the boundary condition $\phi_{\rm even}(\sqrt{\omega}L_y/2,\tilde\varepsilon_n)=0$, then we
will get
\begin{align}\label{even}
M\left(-i\frac{\tilde\varepsilon_n}{2}+\frac{1}{4},\frac{1}{2},i \frac{\omega L^2_y}{8}\right)=0.
\end{align}
Otherwise, if the odd wave function becomes zero at the boundary, $\phi_{\rm odd}(\sqrt{\omega}L_y/2,\tilde\varepsilon_n)=0$, we will thus have
\begin{align}\label{odd}
M\left(-i\frac{\tilde\varepsilon_n}{2}+\frac{3}{4},\frac{3}{2},i \frac{\omega L^2_y}{8}\right)=0.
\end{align}
Notice that the solutions of the above eigenvalue problem depend on both the microscopic details of the electronic band structure, which enter in the definition of the parameter $\omega$, and the ribbon width $L_y$ in the form $\tilde\varepsilon_n(\omega L_y^2)$. Once the set of solutions $\tilde\varepsilon_n$ are obtained, the corresponding energy levels at $q_x=0$ in the conduction band can be evaluated using Eq. (\ref{eq:z-vare}). The energy levels are given by 
\begin{align}
E_c(n,q_x=0)=E_{+}-\frac{\hbar^2}{2m_0}\sqrt{|w^{+}_1(\alpha+\beta+\gamma)|}\tilde\varepsilon_n(\omega L_y^2)
\end{align}
We numerically check that the lowest band has even symmetry, as expected for a symmetric DQW potential.
If we take $q_x\neq0$ then the conduction band Hamiltonian will not commute with parity symmetry, since the potential is asymmetric for finite $q_x$. For any finite $q_x$, we solve Eq. (\ref{eq:boundary}) numerically for the lowest five energy bands and result is shown in Fig. \ref{fig4}(b) which is consistent with the full numerical calculation of the coupled differential equation (see Fig. \ref{fig3}).
This agreement proves that our single-band picture, which predicts the HO and DQW physics in the valence and conduction bands, respectively, are appropriate models for the low-energy physics of strained TMDs. Furthermore, we emphasize that the theory presented here is general and can be straightforwardly adapted to other families of semiconducting 2D crystals.

\subsection{Effect of the scalar potential}\label{Sec:Scalar}

\begin{figure}
\centering
\includegraphics[width=0.7\linewidth]{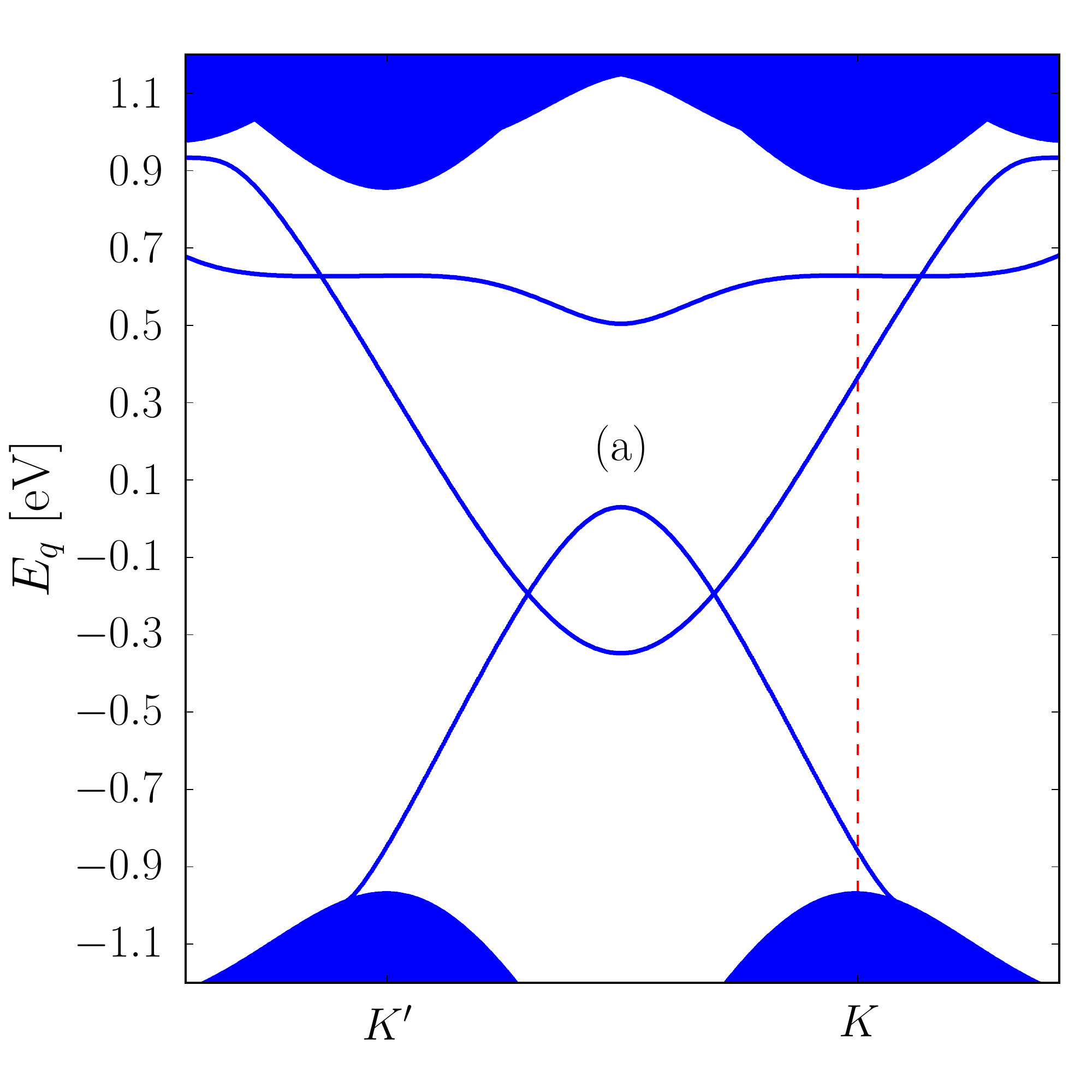}
\includegraphics[width=0.7\linewidth]{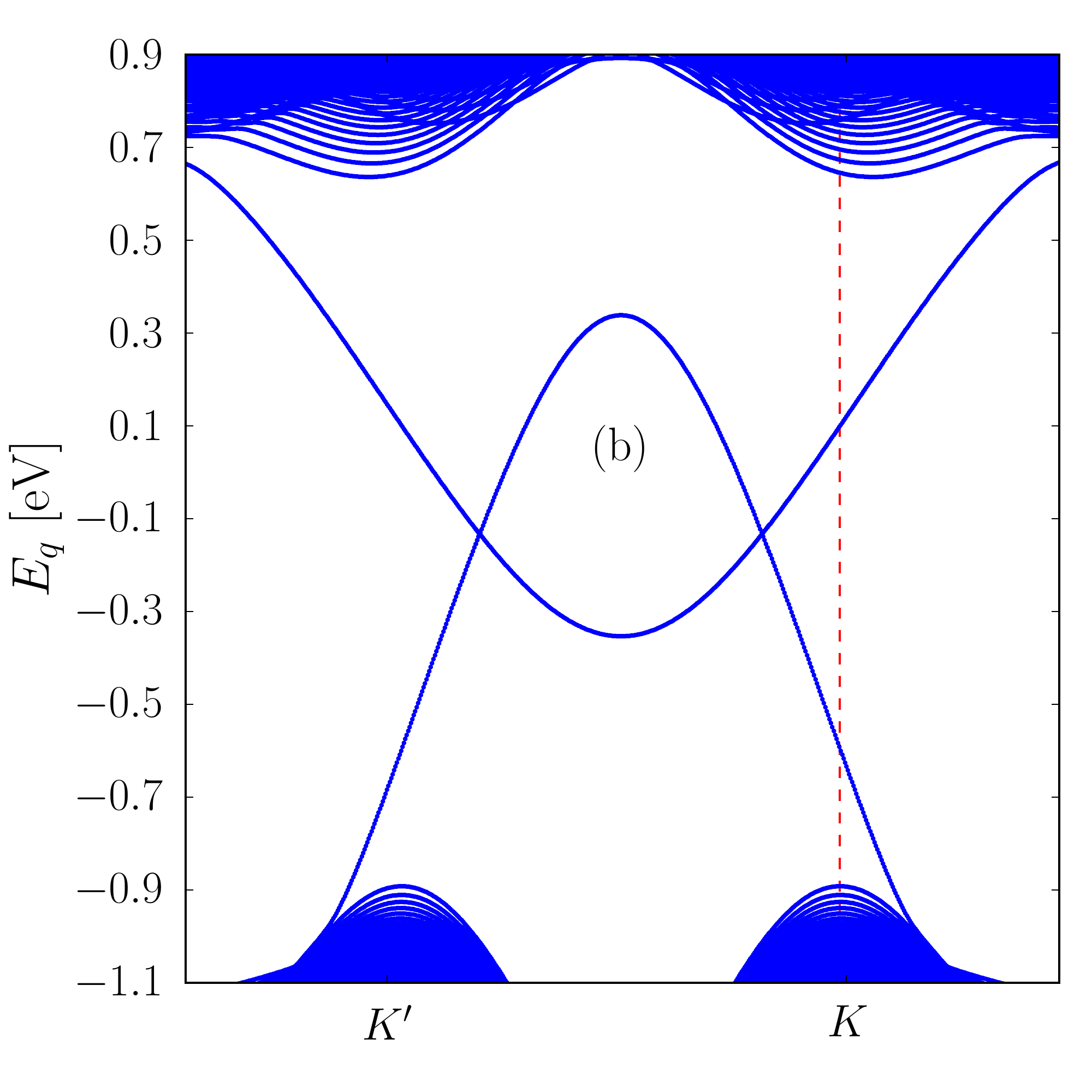}
\includegraphics[width=0.7\linewidth]{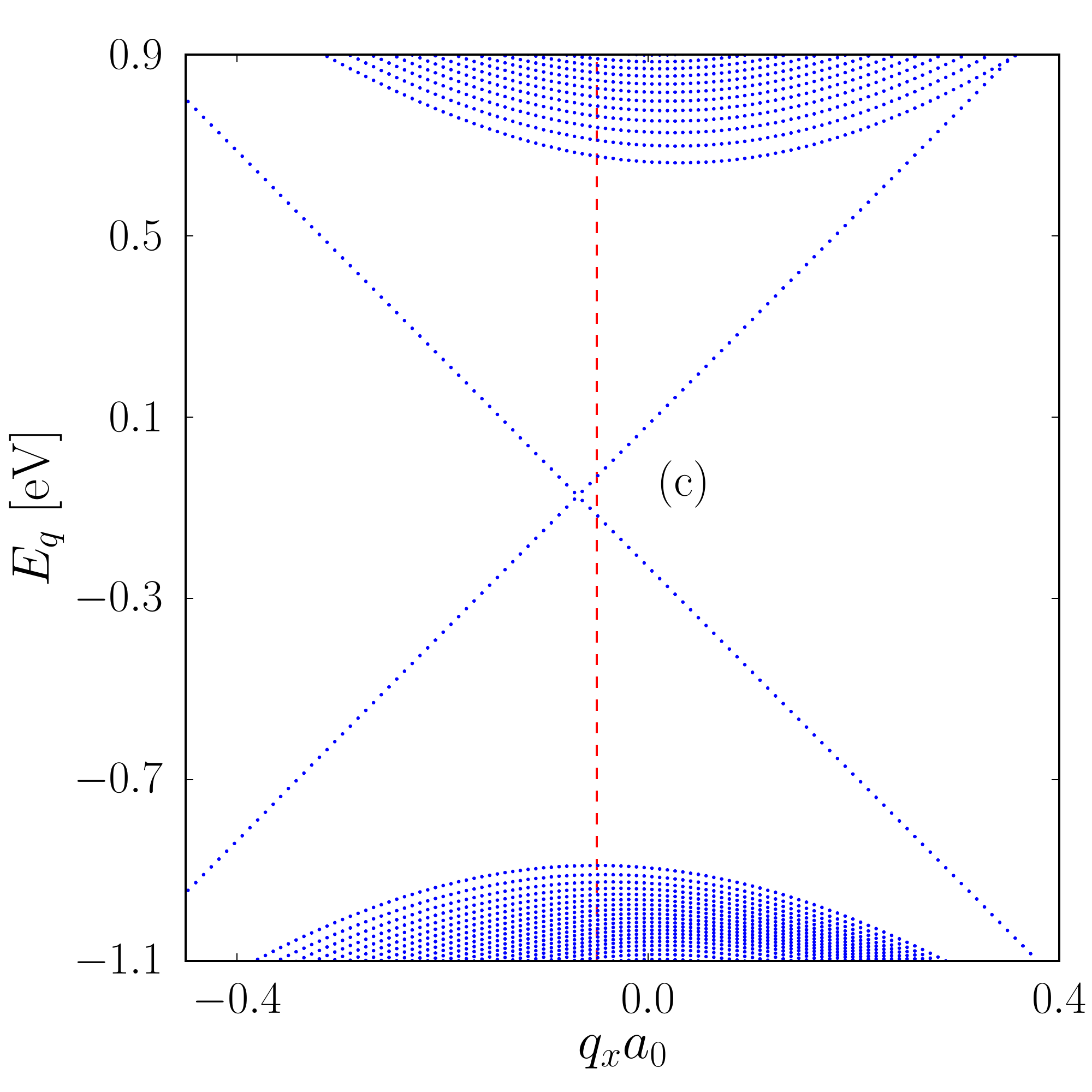}
\caption{(Color online) Energy dispersion in the presence of gauge fields and deformation potentials. (a) Full tight binding calculation in an unstrained zigzag ribbon with ($L_y=299a_0$). (b) Full tight binding calculation in a deformed zigzag ribbon with ($L_y=299a_0$ , $R=7L_y$). Dashed red line indicates a small shift of conduction band minimum with respect to the maximum of the valence band at the K-point. The energy
gap around $K$-point is no longer direct originating from the scalar
potential associated with this profile of the strain. The interlevel
spacing in the conduction and valence bands for
the strained system (panel b) dramatically increases as
compared to the unstrained case (panel a), indicating
that the origin of these levels does not depend on the finite-size effects. (c) Band structure calculated from the low-energy model (\ref{Eq:hks}) on a ribbon with hard wall boundary condition with ($L_y=299a_0$, $R=7L_y$). Notice spin-orbit coupling has not been considered in this figure.
}
\label{fig5}
\end{figure}

In this section, we consider the effect of the scalar potentials on the conduction and valence bands.  We start by performing full TB calculations in a zig-zag ribbon of monolayer MoS$_2$ and the results are shown in Fig. \ref{fig5}(a) and (b) for the unstrained and strained cases, respectively. These results are compared with those results obtained from the low-energy models given by Eq.~(\ref{Eq:hks}). First, one notices the existence of three edge modes in the spectrum, in agreement with previous results~\cite{RA14,RA15}. In Fig.\ref{fig5} (b) we show the results for the band dispersion in the arc-shaped strained case, which contains a set of roughly parabolic valence and conduction bands. Interestingly, the energy gap around $K$-point is no longer direct, which is one of the interesting consequences originating from the scalar potential associated with this profile of the strain.  Moreover, the crossing edge modes survive to the application of strain, whereas the flat high energy edge mode eventually enters into the bulk spectrum for a higher strength of the strain. Intriguingly, one can see that the inter-level spacing in the conduction and valence bands for the strained system (Fig. \ref{fig5}(b)) increases dramatically as compared to the unstrained case (Fig. \ref{fig5}(a)), indicating that the origin of these levels does not depend on the finite size effects but instead of the bulk potential induced by strain. In fact, they are the eigenvalues of the inverted and ordinary harmonic oscillator Hamiltonian of the electrons in the conduction and valence bands, respectively. Finally, Fig. \ref{fig5}(c) shows the results from the two-band low-energy model (\ref{Eq:hks}) for the same system, which are in good agreement with the corresponding results from the full tight-binding model (Fig. \ref{fig5}(b)).

In order to have some analytical insights of the effect of the scalar potential in the band structure of Fig.~\ref{fig5}, we notice that such scalar potential for this strain profile can be written as
\begin{align}\label{eq:deform-pot}
&D_{\pm}(y)=\frac{\hbar^2}{4m_0}[2\kappa^\pm_1 y+\kappa^\pm_2 y^2]\nonumber\\
&\kappa^\pm_1=\frac{4m_0}{\hbar^2}\frac{\alpha^\pm_2}{a_0^2R}\nonumber\\
&\kappa^\pm_2=\frac{8m_0}{\hbar^2}\frac{\alpha^\pm_1+\alpha^\pm_3}{a_0^2R^2}.
\end{align}

To include these potentials in the analytical calculation based on Eq. (\ref{eq:arc-perturb}), we do need to replace $w^\pm_i$ with $v^\pm_i$, where
\begin{align}
&v_1^{\pm}=w_1^{\pm}+\kappa^\pm_2\nonumber\\
&v_2^{\pm}=\frac{w_1^{\pm}w_2^{\pm}-\kappa^\pm_1q_x^{-1}}{v_1^{\pm}}\nonumber\\
&v_3^{\pm}=w_3^{\pm}+w^{\pm}_1w^{\pm2}_2-v_1^{\pm}v_2^{\pm2}.
\end{align}
Using the above relations, the energy dispersion in the valence band can be easily calculated. According to the negative sign of $\alpha^{\pm}_{2}$, one can see that $v^\pm_1$ are negative for any value of strain, which means that the DQW and HO physics in the conduction and valence bands, as discussed above, are still valid for TMDs with arc-shaped deformation in the presence of the scalar potential.

Intriguingly, the shift of the conduction band minimum from the $K$-point should originate from the $\kappa^{+}_1 /q_x$ term in $v^{+}_2$ according to the boundary condition equation given by Eq.~(\ref{eq:boundary}).
In particular, the band edge energy under arc-shaped strain can be expressed as
\begin{align}
E_{{\rm VBM}}&=\frac{\Delta_0-\Delta}{2}+\lambda_{-}-\eta_1\frac{t_0^2}{\Delta-\lambda_{-}}\frac{a_0^2}{l_B^2}
\nonumber\\&
-\frac{\hbar^2}{4m_0}\left[\sqrt{v^{-}_1(\alpha-\beta-\gamma)}-\frac{(\kappa^{-}_1)^2}{v^{-}_1}\right]
\end{align}
The last term proportional to $\kappa^{-}_1$ is independent of strain strength. The strain-independent term should originate from our approximations.

\section{Strain induced Valley shift in homogeneous deformations}\label{Sec:Valley}

\begin{figure}
\centering
\includegraphics[width=0.8\linewidth]{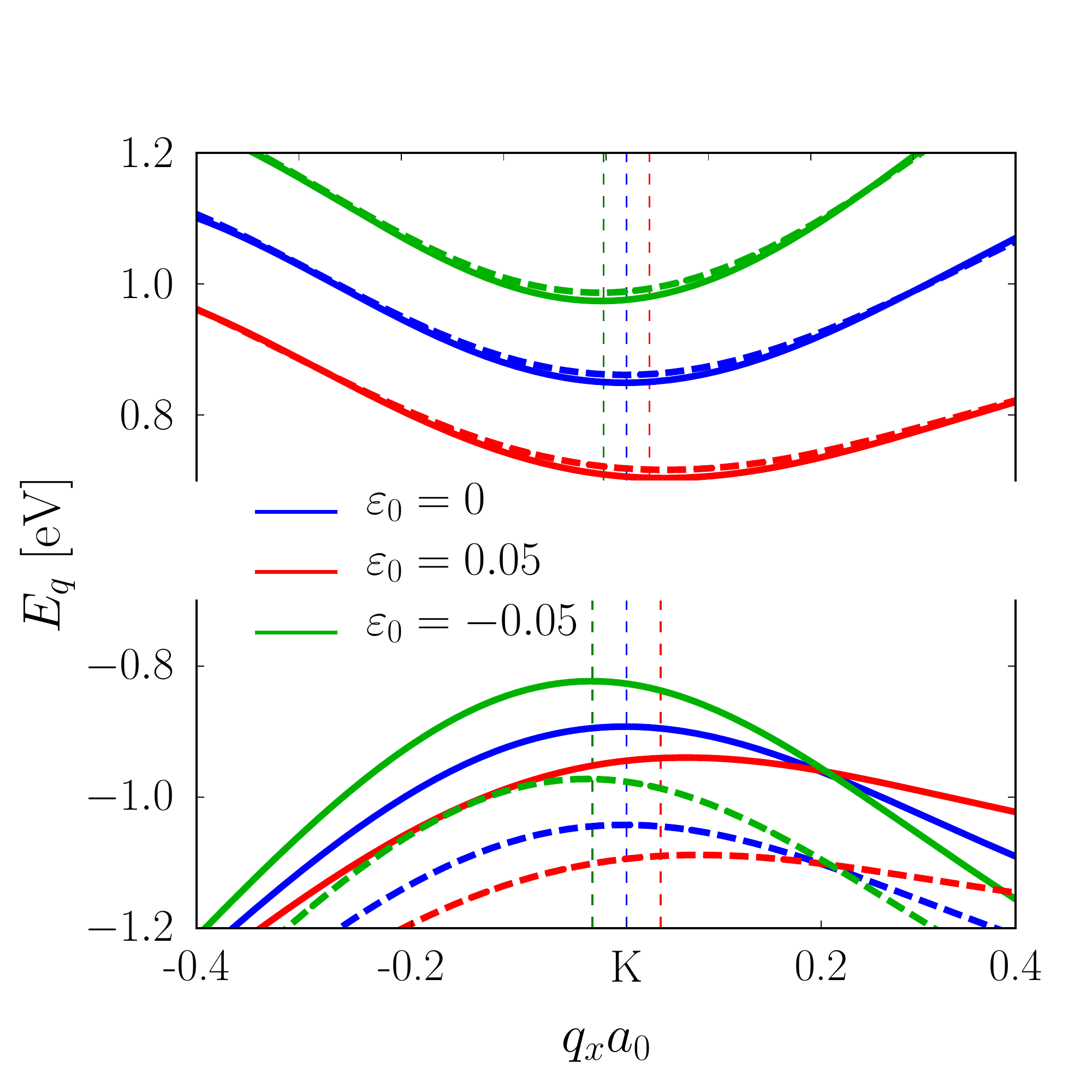}
\caption{(Color online).
Energy dispersion  of a uniaxially deformed
monolayer MoS$_2$
calculated by using the full TB model.
Solid (dashed) lines indicate spin up (down) components.
The vertical dashed lines mark
in addition the position of the conduction band minimum and valence band maximum according to the low-energy model (\ref{v-shc})-(\ref{v-shv}).}
\label{fig6}
\end{figure}

Using the strain induced modification of the hoppings given by Eq. (\ref{Eq:hopping}), one can calculate the band dispersion in the
presence of different kinds of strain profiles.
A teachful example is the case of a uniform uniaxial strain.
It is commonly known, from DFT calculations~\cite{ZS13}, that uniaxial and shear strain induces a shift of the band edges from
the $K$ points, similar to the strained graphene.
We can address this issue in our tight-binding approach.
The energy dispersion of the monolayer MoS$_2$, uniaxially deformed along $x$,
is shown in Fig. \ref{fig6} close to the $K$-
point~\cite{noteUNI}.
Here, solid (dashed) bands indicate spin up (down) components. Both (conduction and valence) band edges shift in phase towards the $\Gamma$ point for compressive ($\varepsilon_0< 0$) strain, whereas they move in the opposite direction for tensile strain, in agreement with DFT simulations~\cite{ZS13}. A useful quantification of these valley-shifts with the strain can be obtained by means of an effective low-energy model.  In the case of uniaxial strain applied along $x$-direction, we have $a_0e/\hbar{\bf A}_1=\eta_1 {\cal A}\hat{x}$ ,$a_0e/\hbar{\bf A}_2=\eta_2 {\cal A}\hat{x}$, and $a_0e/\hbar{\bf A}_3=\eta_3 {\cal A}\hat{x}$ where ${\cal A}=(1-\nu)\varepsilon_0$, where $\nu$ is the Poisson ratio. This uniaxial strain shifts the valleys along the $x$-direction, hence we safely set $q_y=0$. In this case, the approximated Hamiltonian for spin up around the $K$-point is
\begin{align}
H&=\frac{\Delta}{2}\sigma_z+\frac{\lambda_{-}}{2}(1-\sigma_z)+{\cal D}+t_0k_1\sigma_x\nonumber\\&+\frac{\hbar^2}{4m_0a_0^2}(\alpha k_2^2+\beta\sigma_z k_3^2),
\end{align}
where $k_i=q+\eta_i{\cal A}$ and $q=a_0 q_x$. Notice that just the leading term of the spin-orbit coupling is taken into account. The Hamiltonian can be easily diagonalized to obtain its band dispersion. Thus, it is straightforward to find the position of the conduction band minimum ($q_{\rm CBM}$) and  the valence band maximum $(q_{\rm VBM})$, which are given by
\begin{eqnarray}
q_{\rm CBM}
&=&
-{\cal A}\frac{\alpha\eta_2+\beta\eta_3+\gamma\eta_1}{\alpha+\beta+\gamma},
\label{v-shc}
\\
q_{\rm VBM}
&=&
-{\cal A}\frac{\alpha\eta_2-\beta\eta_3-\gamma\eta_1}{\alpha-\beta-\gamma},
\label{v-shv}
\end{eqnarray}
where the  scalar potentials have no contribution to the leading term of the valley-shift. Importantly, in the particle and hole bands, which they have the same effective mass ($\alpha=0$), $q_{\rm VBM}=q_{\rm CBM}=-{\cal A}$ and the position of the valence and conduction band extreme are equally modified by strain. A similar behavior is obtained when $\eta_1=\eta_2=\eta_3=\eta$, in which $q_{\rm VBM}=q_{\rm CBM}=-\eta{\cal A}$. Since none of the previous special conditions apply to the case of strained MoS$_2$, we expect different shifts for the electron and hole band edges. Indeed, based on the numerical value of the parameters in the low-energy model, we find $q_{\rm VBM}=0.76{\cal A}$ and $q_{\rm CBM}=0.51{\cal A}$. Such a different strain induced a shift of the band edges leads to a direct-to-indirect gap transition in MoS$_2$ under uniaxial strain.

It is interesting to compare the above results from the low-energy model with those results obtained from the TB. We do so in Fig.  \ref{fig6}, where the vertical dashed lines indicate the position of the conduction band minimum and valence band maximum as obtained from the low-energy model (\ref{v-shc})-(\ref{v-shv}). In the case of compressive strain, there is a good quantitative agreement between the two methods. In the case of tensile strain, although the qualitative behavior is well captured by the low-energy model, the position of the valence band edge differs in the two cases. The reason is that, according to TB results, tensile strain enhances trigonal warping of the valence band, which is not considered within the simple low-energy model. A similar analysis can be done to understand valley-shifting induced by shear strain. In this case the result is similar to the one for uniaxial strain, with the difference that ${\cal A}=-2\varepsilon_0$ and the deformation is along the $y$-direction. Finally, we remember that for biaxial strain, since ${\cal A}=0$, thus there is no valley-shift.

\section{Spin-strain coupling}\label{Sec:Spin}
Another interesting effect which is worth to be addressed is the direct coupling between the spin and strain. Due to the spin-orbit coupling, we can manipulate spin degree of freedom of carriers just by controlling their orbital motion. Using the direct spin-strain coupling, one can locally control the spin of carriers via the mechanical probe. In the absence of the spin-orbit coupling, there is no way to touch spin degree of freedom by deformation of the lattice via mechanical probe such as AFM tip.\par

Our effective low-energy model (\ref{Eq:hks}) shows that the spin-orbit coupling terms are affected by external strain. At ${\bf q}=0$ the spin-orbit coupling in the conduction and valence bands are modified as follows
\begin{align}
\Delta\lambda_{\pm}&=[\alpha^{s\pm}_1+\lambda'_0|\eta_4|^2\pm\lambda'|\eta_5|^2]|{\cal A}|^2
+\alpha^{s\pm}_2(V+\omega^2_{xy})
\nonumber\\&
+\alpha^{s\pm}_3V^2.
\end{align}
These terms give rise a direct coupling between the spin and strain at the $K$-point. This coupling can allow for spin relaxation when translational symmetry in the system is broken because of, e.g., the existence of ripples which act as a long-range disorder potential. In fact, for the most studied case of graphene, such disorders lead to a spatially random spin-orbit coupling that might have implications for the spin relaxation~\cite{WZ10}. The effect of out-of-plane and in-plane deformation on the spin relaxation in the systems with similar symmetry have been studied~\cite{OGF13,YC15}. Although this effective spin-strain coupling seems to be weak, it is still comparable with some other energy scales like Zeeman energy or weak spin-splitting of the conduction band in semiconducting TMDs. This effect is expected to be especially relevant for the W families of TMDs (like WS$_2$ or WSe$_2$), for which the spin-orbit splitting of the valence band is more than twice the value for MoS$_2$.  From the experimental point of view, this tunable spin-orbit coupling via strain can be detected, in principle, using a photoluminescence measurement~\cite{MH10} because in the strained sample, we expect a change in the relative position of A and B exciton peaks as compared with their position in the undeformed case.

Finally, we notice that for finite value of ${\bf q}$, another spin-dependent term $2a_0e/\hbar\{\lambda'_0{\bf q}.{\bf A}_4+\lambda'{\bf q}.{\bf A}_5\sigma_z\}\otimes s$ appears in the presence of strain. This term is important when ${\cal A}$ is finite and $q$ is small enough ($|{\cal A}|>|a_0{\bf q}|$), because in this regime we have $|a_0{\bf q}||{\cal A}|>|a_0{\bf q}|^2$. Therefore, this new momentum dependent term in (\ref{Eq:hks}) originating from strain (which is $\propto q$) is dominant as compared to the $q$-dependent term in the unstrained case,  which is $\propto q^2$.

\section{summary}\label{Sec:Summary}
In summary, we have studied the strain dependence of monolayer TMDs band structure, starting from a Slater-Koster tight-binding method which contains the necessary orbital contribution to describe the valence and conduction bands in the whole BZ. For a general inhomogeneous strain profile, we further calculate a low-energy Hamiltonian up to second order in momentum, strain and rotation tensors. Our numerical and analytical calculations, based on TB and continuum models, show a strong particle-hole asymmetry in the energy spectrum of the system. We have shown that the momentum dependent terms in the low-energy model of monolayer TMDs acquire different  strain induced vector potentials. According to our calculations using the low-energy model, the electronic spectrum of deformed single layer TMDs reveals a {\it gauge-dependance} which implies that these strain induced vector fields cannot be referred as gauge fields. Consequently, the simple pseudomagnetic field picture like, which is well-known in the case of strained graphene~\cite{VKG10} is no longer valid in deformed monolayer TMDs.

We have applied our theory to calculate the band structure in the illustrative case of arc-shaped deformation. We show that this profile of strain induces three main fictitious gauge fields in Landau's gauge. The dispersion relation of the system within this gauge shows several band crossings in the electron sector of the spectrum, and equidistant parabolic subbands in the hole spectrum. Our analytical calculations show that the energy spectrum in the conduction and valence bands originate, respectively, from the solutions for a double quantum well and a harmonic oscillator Hamiltonian. This DQW and HO physics in two bands is also expected for a triangular strain profile.

Finally, we study the shift of the conduction and valence band edges of $MX_2$ in the presence of homogeneous strain, finding a transition from direct to indirect gap. Moreover, the coupling between spin degrees of freedom and strain has been analyzed. We show that this effect can be considered as a correction on the spin-orbit coupling that can useful for strain engineered spintronic applications.
\acknowledgments
H.R., R.R. and F.G. acknowledge the European Commission under the Graphene Flagship, contract CNECT-ICT-604391. As well as, R.R. and F.G. thank the European Research Council Advanced Grant program (contract 290846). 
E.C. acknowledges support from the European project FP7-PEOPLE-2013-CIG ``LSIE 2D" and from Italian National MIUR Prin
 project 20105ZZTSE. H.R. and E.C. were also supported by MIUR (Italy) through the program ``Progetti Premiali 2012'' - Project ``ABNANOTECH". RR and FG acknowledge FCT-Portugal through grant no. EXPL/FIS-NAN/1728/2013.
\appendix

\section{On-site and hopping matrices}\label{App:TB}

In this Appendix, we provide the analytical expressions for the different contributions to the tight-binding Hamiltonian (\ref{Eq:H-k}). The on-site terms of the Hamiltonian can be written in a compact form~\cite{RO14}:
\begin{eqnarray}
\boldsymbol{\epsilon}
&=&
\left(
\begin{array}{cc}
\epsilon_M & 0 \\
0 & \epsilon_X
\end{array}
\right),
\end{eqnarray}
where
\begin{eqnarray}
\epsilon_M
&=&\begin{pmatrix}\epsilon_0&&0&&0\\0&&\epsilon_2&&-i\lambda_M\hat{s}_z\\0&&i\lambda_M\hat{s}_z&&\epsilon_2\end{pmatrix},
\nonumber\\
\nonumber\\
\epsilon_X
&=&\begin{pmatrix}\epsilon_p+t^\perp_{xx}&&-i\frac{\lambda_X}{2}\hat{s}_z&&0\\i\frac{\lambda_X}{2}\hat{s}_z&&\epsilon_p+t^\perp_{yy}&&0\\0&&0&&\epsilon_z-t^\perp_{zz}\end{pmatrix}.
\end{eqnarray}
Here, $\lambda_M$ and $\lambda_X$ stand for the spin-orbit coupling of $M$ (metal) and $X$ (chalcogen) atoms, respectively \cite{RO14} and $\hat{s}_z=\pm$ indicates $z$-component of spin degree of freedom.
The terms $t^\perp_{xx}=t^\perp_{yy}=V_{pp\pi}$, $t^\perp_{zz}=V_{pp\sigma}$
take into account the effects of the vertical hopping $V_{pp}$
between the top and bottom chalcogen atoms.

Below we list the hopping terms of the model. For the nearest neighbor hopping between $M$ and $X$ atoms we have~\cite{CS13}
\begin{widetext}
\begin{align}
t^{MX}_1&=\frac{\sqrt{2}}{7\sqrt{7}}\begin{pmatrix}-9 V_{pd\pi}+\sqrt{3}V_{pd\sigma}&&3\sqrt{3}V_{pd\pi}-V_{pd\sigma}&&12 V_{pd\pi}+\sqrt{3} V_{pd\sigma}\\
          5\sqrt{3} V_{pd\pi}+3 V_{pd\sigma}&&9 V_{pd\pi}-\sqrt{3} V_{pd\sigma}&&-2\sqrt{3}V_{pd\pi}+3 V_{pd\sigma}\\
          -V_{pd\pi}-3\sqrt{3}V_{pd\sigma}&&5\sqrt{3}V_{pd\pi}+3 V_{pd\sigma}&&6 V_{pd\pi}-3\sqrt{3} V_{pd\sigma}\end{pmatrix}\\
\nonumber\\
t^{MX}_2&=\frac{\sqrt{2}}{7\sqrt{7}}\begin{pmatrix}0&&-6\sqrt{3}V_{pd\pi}+2V_{pd\sigma}&&12V_{pd\pi}+\sqrt{3}V_{pd\sigma}\\
          0&&-6V_{pd\pi}-4\sqrt{3}V_{pd\sigma}&&4\sqrt{3} V_{pd\pi}-6V_{pd\sigma}\\14V_{pd\pi}&&0&&0\end{pmatrix}\\
\nonumber\\
t^{MX}_3&=\frac{\sqrt{2}}{7\sqrt{7}}\begin{pmatrix}9 V_{pd\pi}-\sqrt{3}V_{pd\sigma}&&3\sqrt{3}V_{pd\pi}-V_{pd\sigma}&&12 V_{pd\pi}+\sqrt{3} V_{pd\sigma}\\
          -5\sqrt{3} V_{pd\pi}-3 V_{pd\sigma}&&9 V_{pd\pi}-\sqrt{3} V_{pd\sigma}&&-2\sqrt{3}V_{pd\pi}+3 V_{pd\sigma}\\
          -V_{pd\pi}-3\sqrt{3}V_{pd\sigma}&&-5\sqrt{3}V_{pd\pi}-3 V_{pd\sigma}&&-6V_{pd\pi}+3\sqrt{3} V_{pd\sigma}\end{pmatrix}
\end{align}
\end{widetext}

Next nearest neighbor hoppings correspond to processes between the same kind of atoms, $M$-$M$ or $X$-$X$ (see Fig. \ref{Fig:Lattice}), and they are given by
\begin{widetext}
\begin{align}
t^{MM}_1&=\frac{1}{4}\begin{pmatrix}3V_{dd\delta}+V_{dd\sigma}&&\frac{\sqrt{3}}{2}(-V_{dd\delta}+V_{dd\sigma})&&-\frac{3}{2}(V_{dd\delta}-V_{dd\sigma})\\
        \frac{\sqrt{3}}{2}(-V_{dd\delta}+V_{dd\sigma})&&\frac{1}{4}(V_{dd\delta}+12V_{dd\pi}+3V_{dd\sigma})&&\frac{\sqrt{3}}{4}(V_{dd\delta}-4V_{dd\pi}+3V_{dd\sigma})\\
        -\frac{3}{2}(V_{dd\delta}-V_{dd\sigma})&&\frac{\sqrt{3}}{4}(V_{dd\delta}-4V_{dd\pi}+3V_{dd\sigma})&&\frac{1}{4}(3V_{dd\delta}+4V_{dd\pi}+9V_{dd\sigma})\end{pmatrix}\\
\nonumber\\
t^{MM}_2&=\frac{1}{4}\begin{pmatrix}3V_{dd\delta}+V_{dd\sigma}&&\sqrt{3}(V_{dd\delta}-V_{dd\sigma})&&0\\\sqrt{3}(V_{dd\delta}-V_{dd\sigma})&&V_{dd\delta}+3V_{dd\sigma}&&0\\0&&0&&4V_{dd\pi}\end{pmatrix}\\
\nonumber\\
t^{MM}_3&=\frac{1}{4}\begin{pmatrix}3V_{dd\delta}+V_{dd\sigma}&&\frac{\sqrt{3}}{2}(-V_{dd\delta}+V_{dd\sigma})&&\frac{3}{2}(V_{dd\delta}-V_{dd\sigma})\\
        \frac{\sqrt{3}}{2}(-V_{dd\delta}+V_{dd\sigma})&&\frac{1}{4}(V_{dd\delta}+12V_{dd\pi}+3V_{dd\sigma})&&-\frac{\sqrt{3}}{4}(V_{dd\delta}-4V_{dd\pi}+3V_{dd\sigma})\\
        \frac{3}{2}(V_{dd\delta}-V_{dd\sigma})&&-\frac{\sqrt{3}}{4}(V_{dd\delta}-4V_{dd\pi}+3V_{dd\sigma})&&\frac{1}{4}(3V_{dd\delta}+4V_{dd\pi}+9V_{dd\sigma})\end{pmatrix}\\
\nonumber\\
t^{XX}_1&=\frac{1}{4}\begin{pmatrix}3V_{pp\pi}+V_{pp\sigma}&&\sqrt{3}(V_{pp\pi}-V_{pp\sigma})&&0\\
                                     \sqrt{3}(V_{pp\pi}-V_{pp\sigma})&&V_{pp\pi}+3V_{pp\sigma}&&
                                     0\\0&&0&&4V_{pp\pi}\end{pmatrix}\\
\nonumber\\
t^{XX}_2&=\begin{pmatrix}V_{pp\sigma}&&0&&0\\0&&V_{pp\pi}&&0\\0&&0&&V_{pp\pi}\end{pmatrix}\\
\nonumber\\
t^{XX}_3&=\frac{1}{4}\begin{pmatrix}3V_{pp\pi}+V_{pp\sigma}&&-\sqrt{3}(V_{pp\pi}-V_{pp\sigma})&&0\\-\sqrt{3}(V_{pp\pi}-V_{pp\sigma})&&V_{pp\pi}+3V_{pp\sigma}
                                      &&0\\0&&0&&4V_{pp\pi}\end{pmatrix}.
\end{align}
\end{widetext}
The direction of the hopping indicated by subindexes 1,2, and 3 can be seen in Fig.~\ref{Fig:Lattice}.

\section{Derivation of the low-energy Hamiltonian}\label{App:k.p}
\numberwithin{equation}{section}
In this Appendix, we calculate the low-energy Hamiltonian around the high symmetry $K$ points using L\"{o}wding partitioning method \cite{W03}. The approach is similar to the one used in Ref. [\onlinecite{RMA13}]. Here, the local strain is introduced by means of a local change of the two-center Slater-Koster matrix elements as a consequence of the local modulation of the interatomic bond lengths. We assume a general form of the inhomogeneous deformation with a large wavelength. To consider such deformation, we use the following relations for the bond lengths
\begin{widetext}
\begin{align}\label{bondlength}
&r^{MX}_1=a\sqrt{\left(\frac{1}{2}+\frac{\tilde u_{xx}}{2}-\frac{\tilde u_{yx}}{2\sqrt{3}}\right)^2+\left(\frac{1}{2\sqrt{3}}+\frac{\tilde u_{yy}}{2\sqrt{3}}-\frac{\tilde u_{xy}}{2}\right)^2+\frac{1}{4}}\nonumber\\
&r^{MX}_2=a\sqrt{\left(\frac{\tilde u_{yx}}{\sqrt{3}}\right)^2+\left(\frac{1}{\sqrt{3}}+\frac{\tilde u_{yy}}{\sqrt{3}}\right)^2+\frac{1}{4}}\nonumber\\
&r^{MX}_3=a\sqrt{\left(\frac{1}{2}+\frac{\tilde u_{xx}}{2}+\frac{\tilde u_{yx}}{2\sqrt{3}}\right)^2+\left(\frac{1}{2\sqrt{3}}+\frac{\tilde u_{yy}}{2\sqrt{3}}+\frac{\tilde u_{xy}}{2}\right)^2+\frac{1}{4}}\nonumber\\
&r^{MM(XX)}_1=a\sqrt{\left(\frac{1}{2}+\frac{\tilde u_{xx}}{2}-\frac{\sqrt{3}\tilde u_{yx}}{2}\right)^2+\left(\frac{\sqrt{3}}{2}+\frac{\sqrt{3}\tilde u_{yy}}{2}-\frac{\tilde u_{xy}}{2}\right)^2}\nonumber\\
&r^{MM(XX)}_2=a\sqrt{\left(1+\tilde u_{xx}\right)^2+\tilde u_{xy}^2}\nonumber\\
&r^{MM(XX)}_3=a\sqrt{\left(\frac{1}{2}+\frac{\tilde u_{xx}}{2}+\frac{\sqrt{3}\tilde u_{yx}}{2}\right)^2+\left(\frac{\sqrt{3}}{2}+\frac{\sqrt{3}\tilde u_{yy}}{2}+\frac{\tilde u_{xy}}{2}\right)^2}\nonumber\\
\end{align}
\end{widetext}
in which ${\tilde u}_{ij}=\partial u_j/\partial x_i$.

 To obtain the low-energy model given in Eq.~(\ref{Eq:hks}), we follow the next eight steps:
 \begin{enumerate}
\item We take the six-band tight-binding Hamiltonian (\ref{Eq:H-k}) for a given spin subspace in the deformed system as follows
\begin{align}\label{Eq:H6x6}
H({\bf k},{\bf \tilde u})=H_{\rm TB}[6\times 6]
\end{align}
in which ${\bf \tilde u}$ is a tensor with matrix elements $\tilde u_{ij}$.

\item We expand the Hamiltonian (\ref{Eq:H6x6}) around the $K$ point of the BZ, ${\rm {\bf K}}=4\pi/3a(1,0)$, obtaining
\begin{align}
H({\bf k},{\bf \tilde u})\approx H_{0}+H_1(\xi)
\end{align}
 where $H_{0}=H({\bf K},0)$ and  $\xi=\{q_x,q_y,\tilde u_{xx},\tilde u_{yy},\tilde u_{xy},\tilde u_{yx}\}$, where ${\bf q}={\bf k}-{\bf K}$ and $|{\bf q}|a\ll 1$. $H_1(\xi)$ is obtained as
 \begin{align}
 H_1(\xi)&= \sum_i \xi_i \frac{ \partial H({\bf K}+{\bf q},{\bf \tilde u})}{\partial {\xi_i}}\Big |_{\xi=0}
 \nonumber\\&+\frac{1}{2}\sum_{i} \xi^2_i \frac{ \partial^2 H({\bf K}+{\bf q},{\bf \tilde u})}{\partial {\xi^2_i}}\Big |_{\xi=0}
 \nonumber\\& +\sum_{i\neq j} \xi_i \xi_j \frac{ \partial^2 H({\bf K}+{\bf q},{\bf \tilde u})}{\partial {\xi_i} \partial {\xi_j}}\Big |_{\xi=0}
 \end{align}
In this expansion, we assume that $\xi_i$ and $\xi_j$ commute, which is the case of the homogenous deformation. The generalization for the inhomogeneous strain case can be done, for long wavelength strain profiles, by replacing $\varepsilon_{ij}$, $\omega_{xy}$, and $q_i \varepsilon_{jk}$ with $\varepsilon_{ij}({\bf r})$, $\omega_{xy}({\bf r})$, and $-i(\partial_{r_i} \varepsilon_{jk}({\bf r})+ \varepsilon_{jk}({\bf r})\partial_{r_i})/2$, respectively.

 \item To find the low-energy Hamiltonian in the subspace of the conduction band minimum (CBM) and valence band maximum (VBM), we do a transformation in orbital space using the unitary operator $U_0$, which diagonalizes $H_0$. After solving the eigenvalue problem
\begin{align}
H_0|\psi^i_0\rangle=E^i_0|\psi^i_0\rangle
\end{align}
we obtain
\begin{align}
U_0=(|\psi^1_0\rangle,|\psi^2_0\rangle,|\psi^3_0\rangle,|\psi^4_0\rangle,|\psi^5_0\rangle,|\psi^6_0\rangle)
\end{align}
where we have ordered the eigenstates such that $|\psi^5_0\rangle$ and $|\psi^6_0\rangle$ correspond to the lowest conduction and highest valence band for the given spin, respectively.
Then, we apply this unitary transformation to move from orbital basis ($H$) to the band basis ($H'$) as
\begin{align}
 H'=U_0^\dagger [H_{0}+H_1(\xi)]U_0.
\end{align}

 \item  We do the following replacement in $H'$
\begin{align}
&\tilde u_{xx}=\varepsilon_{xx},\nonumber\\&
\tilde u_{yy}=\varepsilon_{yy},\nonumber\\& \tilde u_{xy}=\varepsilon_{xy}+\omega_{xy}, \nonumber\\&
\tilde u_{yx}=\varepsilon_{xy}-\omega_{xy}.
\end{align}
Now, the Hamiltonian $H'$ is a function of wave vector, ${\bf q}$, symmetric strain tensor, ${\bm \varepsilon}$, and anti-symmetric rotation tensor, ${\bm \omega}$.

\item  We  decompose $H'$ into two parts $H'=H_d+V$
\begin{eqnarray}\label{Eq:Hblocks}
H_d=\begin{pmatrix}h_{11}[4\times4]&&0\\0&&h_{22}[2\times2]\end{pmatrix}\nonumber\\
V=\begin{pmatrix}0&&h_{12}[4\times2]\\h_{12}^\dagger[2\times4]&&0\end{pmatrix}
\end{eqnarray}
where $h_{11}$ is defined in the subspace $\{|\psi^1_0\rangle,|\psi^2_0\rangle,|\psi^3_0\rangle,|\psi^4_0\rangle\}$, whereas $h_{22}$ is defined in the subspace  $\{|\psi^5_0\rangle,|\psi^6_0\rangle\}$. Notice that the high energy $h_{11}$ and low-energy $h_{22}$ blocks in (\ref{Eq:Hblocks}) is coupled with the off-diagonal element $V$. The analytical expression of the block components, i.e. $h_{ij}$, are too lengthy to be included here.

\item An additional unitary rotation is performed to project $V$ into each of these subspaces. We employ the quasi-degenerate perturbation theory by using $e^{-\cal O}$ as rotation operator.
This allows to drop the first-order $V$ in the transformed
Hamiltonian, $H''=e^{-\cal O} H' {e^{\cal O}}=H_d+V+[H_d,{\cal O}]+[V ,{\cal O}]+\frac{1}{2}[[H_d, {\cal
O}], {\cal O}]+\cdots$, leading to the constraint
$V+[H_d,{\cal O}]=0$. The generator of the transformation takes the form
\begin{eqnarray}\label{eq:O}
{\cal O}=\begin{pmatrix}0&\eta[4\times2]\\-\eta^\dagger[2\times4]&0\end{pmatrix},
\end{eqnarray}
where $\eta h_{22}- h_{11}\eta=h_{12}$ is solved to find the $\eta$ matrix as a function of  $\{q_x,q_y, \epsilon_{xx},\epsilon_{yy},\epsilon_{xy},\omega_{xy}\}$ up to second order for the given spin index.

\item Then, $ H''= H_d +\frac{1}{2}[V,{\cal O}]+\cdots$ is our final effective Hamiltonian with two decoupled subspaces. Following a straightforward calculation, the effective Hamiltonian of the low-energy bands can be obtained  for a given spin index as follows,
\begin{align}\label{H2b}
H_{2b}(q_x,q_y, \epsilon_{xx},\epsilon_{yy},\epsilon_{xy},\omega_{xy})&=\nonumber\\
h_{22}+\frac{1}{2}\{\eta^\dagger {h_{12}}+h_{12}^\dagger\eta\}&.
\end{align}

\item Finally, we consider the relations
\begin{align}
{\rm Re} [{\cal A}]&=\varepsilon_{xx}-\varepsilon_{yy},\nonumber\\
{\rm Im} [{\cal A}]&=-2\varepsilon_{xy},\nonumber\\
V&=\varepsilon_{xx}+\varepsilon_{yy}
\end{align}
and factorize (\ref{H2b}) to reach the form of the Hamiltonian Eq. (\ref{Eq:hks}). Then, we simply extract the numerical value
 of all parameters in our low-energy model i.e. $\{t_0,t_1,t_2,\alpha,\beta, \lambda_0,\lambda,\Delta_0,\Delta,
\alpha',\beta',\lambda'_0,\lambda',
\eta_1,\eta_2,\eta_3,\eta_4,\eta_5,
\\
\alpha^\pm_1,\alpha^\pm_2,\alpha^\pm_3,\alpha^{s\pm}_1,\alpha^{s\pm}_2,\alpha^{s\pm}_3\}$
from the numerical values of the pre-factors, and the result is given in Table \ref{Tab:t_low}.
\end{enumerate}

We emphasize that the extension to the inhomogeneous case is already done in (\ref{Eq:hks}) just by considering the local nature of the spin-independent contribution to the scalar potential (${\cal D}$), and its spin-dependent part (${\cal \delta\lambda}$), as well as non-zero commutation of the momentum and the fictitious vector fields (i.e. $[{\bf q},{\bf A}_i]\neq 0$ ). This kind of extension from homogeneous to inhomogeneous deformation is common in studying strained conventional semiconductors and graphene \cite{Z94,L12}.


Note that there is also a {\it trigonal warping} term which is not included in Hamiltonian (\ref{Eq:hks}). For the unstrained case, it has the form
\begin{align}
H_w=t_1 a_0^2{\bf q}\cdot{\sigma}^{\ast}_{\tau}\sigma_x {\bf q}\cdot{\sigma}^{\ast}_{\tau}+t_2a_0^3\tau(q_x^3-3q_xq_y^2)(\alpha'+\beta'\sigma_z)
\end{align}
where $t_1=-0.14$eV, $t_2=1$eV, $\alpha'=0.44$, $\beta'=-0.53$. Here, the trigonal warping term contains three parameters ($\alpha',\beta',t_1$). It is easy to show that all these terms combine with each other to lead the characteristic contribution $z_{\pm}\cos3\phi$ to the low-energy dispersion at the K-point, where $z_{\pm}=t_2(\alpha'\pm\beta')\pm2 t_0 t_1/(\Delta-\lambda_{-}\tau s)$, and $z_+(z_-)$ stands for the conduction (valence) band.

It is interesting to notice that the direction of warping in both bands is opposite if
$z_{+} z_{-}>0$, and it is the same otherwise. If $\alpha'=0$ the warping in both bands are in the same direction and  with same warping strength.  Furthermore, $\alpha$ and $\alpha'$ are the sources of asymmetry in effective masses and trigonal warping directions between the conduction and the valence band, respectively. In our case, $z_{+} z_{-}<0$ which means same warping direction in the two bands.

\section{Derivation of the single band model}\label{derivation-single-band}
Since ${\bf A}_1$ is weak as compared to the other vector potentials entering in our theory, we can apply perturbation methods to decouple the conduction and the valence bands, driving the two-bands Hamiltonian (\ref{Eq:hks}) into two one-band Hamiltonians as given in Eq. (\ref{eq:arc-perturb}). To do this we use a canonical transformation similar to what we have done to get (\ref{Eq:hks}) in Appendix \ref{App:k.p}, in which $H_d={\rm diag}[h_c,h_v]$ and $V_{12}=V^\dagger_{21}=h_{cv}$.
By neglecting the spin splitting in the conduction band and the momentum dependence of the spin-splitting in the valence band, we obtain an expression equivalent to (\ref{Eq:Hblocks}) by defining the following relations
\begin{align}
&h_c=\frac{\Delta_0+\Delta}{2}+D_{+}+\frac{b a_0^2}{\hbar^2}\alpha{ \pi}_2^2+\frac{b a_0^2}{\hbar^2}\beta{ \pi}_3^2\nonumber\\
&h_v=\frac{\Delta_0-\Delta}{2}+D_{-}+\lambda_{-}+\frac{b a_0^2}{\hbar^2}\alpha{ \pi}_2^2-\frac{b a_0^2}{\hbar^2}\beta{ \pi}_3^2\nonumber\\
&h_{cv}=\frac{t_0 a_0}{\hbar} { \pi}_1^\dagger
\end{align}
where $b=\frac{\hbar^2}{4m_0 a_0^2}$, $h_c=h_{11},h_v=h_{22}$,  and $h_{cv}=h_{12}$.
To calculate the generator of the transformation
\begin{align}
{\cal O}=\begin{pmatrix}0&&\vartheta\\-\vartheta^\dagger&&0\end{pmatrix}
\end{align}
we must first find $\vartheta$, which obeys the following relation
\begin{align}\label{eq:theta}
\vartheta=-h_{cv}(h_c-h_v)^{-1}-[\vartheta,h_c](h_c-h_v)^{-1}
\end{align}
We solve Eq. (\ref{eq:theta}) in an iterative perturbative approach in the low-momentum and low-strain limits. In this regard, it is easy to show that
\begin{align}
\vartheta=\frac{\vartheta^{(1)}}{\Delta-\lambda_{-}}+\frac{\vartheta^{(2)}}{(\Delta-\lambda_{-})^2}+\dots
\end{align}
where
\begin{align}
&\vartheta^{(1)}=-\frac{t_0a_0}{\hbar} { \pi}_1^\dagger\nonumber\\
&\vartheta^{(2)}=2\frac{t_0a_0}{\hbar}\{[{ \pi}_1^\dagger,h_c]+\frac{b a_0^2}{\hbar^2}\beta{ \pi}_1^\dagger{ \pi}_3^2\}.
\end{align}
Therefore, performing the canonical transformation on the two-band Hamiltonian (\ref{Eq:hks}) we obtain, to first order in $h_{cv}$, two decoupled Hamiltonians for the conduction and valence bands, $H_{c(v)}=H_{c(v)}^{(1)}+H_{c(v)}^{(2)}$, where
\begin{align}
H^{(1)}_{c}&=\frac{\Delta_0+\Delta}{2}+D_{+}+\eta_1\frac{t_0^2}{\Delta-\lambda_{-}}\frac{a_0^2}{l_B^2}
\nonumber\\&
+\frac{\hbar^2}{4m_0}(\alpha|{\bf q}+\frac{e}{\hbar}{\bf A}_2|^2
+\beta|{\bf q}+\frac{e}{\hbar}{\bf A}_3|^2+\gamma|{\bf q}+\frac{e}{\hbar}{\bf A}_1|^2)\nonumber\\
\end{align}
\begin{align}
H^{(1)}_{v}&=\frac{\Delta_0-\Delta}{2}+D_{-}+\lambda_{-}-\eta_1\frac{t_0^2}{\Delta-\lambda_{-}}\frac{a_0^2}{l_B^2}\nonumber\\&+\frac{\hbar^2}{4m_0}(\alpha|{\bf q}+\frac{e}{\hbar}{\bf A}_2|^2
-\beta|{\bf q}+\frac{e}{\hbar}{\bf A}_3|^2-\gamma|{\bf q}+\frac{e}{\hbar}{\bf A}_1|^2)\nonumber\\
\end{align}
where $\gamma=4m_0 v^2/(\Delta-\lambda_{-})$. We can also consider higher second order terms originating from the expansion (\ref{eq:theta}), with the result
\begin{align}
H^{(2)}_{c}&=\frac{1}{(\Delta-\lambda_{-})^2}\frac{t_0^2a_0^2}{\hbar^2}\frac{b a_0^2}{\hbar^2} \{
\frac{\hbar^2}{l_B^2}[\alpha(\eta_1+\eta_2)({ \pi}_1^\dagger{ \pi}_2+{ \pi}_2^\dagger{ \pi}_1)\nonumber\\&+\beta(\eta_1+\eta_3)({ \pi}_1^\dagger{ \pi}_3+{ \pi}_3^\dagger{ \pi}_1)]
-2\beta { \pi}_1^\dagger { \pi}_3^2 { \pi}_1 \}
\end{align}
for the conduction band, and
\begin{align}
H^{(2)}_{v}&=\frac{1}{(\Delta-\lambda_{-})^2}\frac{t_0^2a_0^2}{\hbar^2}\frac{b a_0^2}{\hbar^2} \{
\frac{\hbar^2}{l_B^2}[\alpha(\eta_1+\eta_2)({ \pi}_1{ \pi}_2^\dagger+{ \pi}_2{ \pi}_1^\dagger)\nonumber\\&+\beta(\eta_1+\eta_3)({ \pi}_1{ \pi}_3^\dagger+{ \pi}_3{ \pi}_1^\dagger)]
+\beta ({\pi}_1 {\pi}_1^\dagger { \pi}_3^2+{ \pi}_3^2 { \pi}_1 { \pi}_1^\dagger ) \}
\end{align}
for the valence band, where we have neglected the contribution from the scalar potential term $D_{\pm}$.

\section{Numerical eigenvalue problem}\label{numerical-fem}
\numberwithin{equation}{section}

To discritize and find the eigenvalues of the Hamiltonian for an arc-shaped monolayer TMDs, we use the method of moments \cite{H93} with a trigonal basis to satisfy the hard wall boundary conditions which require vanishing of the wave function at the boundaries. We can expand the wave function in a set based on trigonal basis $T_n(y)=T(y-y_n)$ as $\phi_c=\sum_{n} \kappa_n T_n(y)$ and $\phi_v=\sum_{n} \chi_n T_n(y)$, where $\phi_c$ and $\phi_v$ refer to the conduction and valence band spinor components respectively, and
\begin{eqnarray}
T(y)=\Big\{
  \begin{array}{l l}
    L_y-|y|(N+1) & \quad \text{$|y|<L_y/(N+1)$}\\
    0 & \quad \text{otherwise}
  \end{array}
\end{eqnarray}
and $y_n=L_y(\frac{n}{N+1}-\frac{1}{2})$. Note that $L_y$ and $N$ are the width of the system along the $y$-direction and the number of basis functions, respectively.
It should be mentioned that using this trigonal basis guarantees the zero value of the wave function at the boundaries.
To do this numerical calculation we need to know the following matrix elements in this set of trigonal basis
\begin{widetext}
\begin{eqnarray}
&&\langle T_m|\partial_y^2|T_n \rangle=-2(N+1)\delta_{m,n}+(N+1)(\delta_{m,n+1}+\delta_{m,n-1})\nonumber\\
&&\langle T_m|\partial_y|T_n \rangle=\frac{L_y^2}{2}(\delta_{m,n+1}-\delta_{m,n-1})\nonumber\\
&&\langle T_m|T_n \rangle=\frac{L_y^3}{(1+N)}\left[\frac{2}{3}\delta_{m,n}+\frac{1}{6}(\delta_{m,n+1}+\delta_{m,n-1})\right]\nonumber\\
&&\langle T_m|y|T_n \rangle=\frac{L_y^4}{(1+N)^2}\left[\frac{2n-1-N}{3}\delta_{m,n}+\frac{2n-N}{12}\delta_{m,n-1}+\frac{2n-2-N}{12}\delta_{m,n-1}\right]\nonumber \\
&&\langle T_m|y^2|T_n \rangle=\frac{L_y^5}{(1+N)^3}\left[\frac{10(n-\frac{N+1}{2})^2+1}{15}\delta_{m,n}+\frac{20(n-\frac{N}{2})^2+1}{120}\delta_{m,n+1}
+\frac{20(n-\frac{N+2}{2})^2+1}{120}\delta_{m,n-1}\right].\nonumber \\
\end{eqnarray}
\end{widetext}
Using these matrix elements one can discretize the Hamiltonian (\ref{hpll2}). Since trigonal basis has non-zero overlap, using this method we obtain a generalized eigenvalue problem that we solve numerically. The results can be seen in Figs. (\ref{fig2}) and (\ref{fig3}).

\bibliography{BibliogrGrafeno}

\begin{thebibliography}{72}
\bibitem [{not()}]{noteUNI}%
  \BibitemOpen
  \href@noop {} {}\bibinfo {note} {We remind that, in order to disentangle pure
  geometrical effects from the role of the lattice distortion on the hopping
  parameters, we model strain uniquely through the modulation of these latter
  ones (hopping parameters), without a geometrical deformation of the lattice.
  In the perspective the Brillouin zone is unaffected in the presence of
  strain.}\BibitemShut {Stop}%
\end{thebibliography}

\end{document}